# Gravitational-Wave research as an Emerging Field in the Max Planck Society.

# The Long Roots of GEO600 and of the Albert Einstein Institute[1]

Luisa Bonolis[2] and Juan-Andres Leon[3]

*Max Planck Institute for the History of Science, Berlin*

*Abstract*

On the occasion of the 50[th] anniversary since the beginning of the search for gravitational waves at the Max Planck Society, and in coincidence with the 25[th] anniversary of the foundation of the Albert Einstein Institute, we explore the interplay between the *renaissance* of general relativity and the advent of relativistic astrophysics following the German early involvement in gravitational-wave research, from the very first interests of its scientists, to the point when gravitational-wave detection became established by the appearance of full-scale detectors and international collaborations.

On the background of the spectacular astrophysical discoveries of the 1960s and the growing role of relativistic astrophysics, Ludwig Biermann and his collaborators at the Max Planck Institute for Astrophysics in Munich became deeply involved in research related to such new horizons, already unveiled by radio astronomy during the 1950s, and further widened by the advent of X-ray and gamma-ray astronomy.

At the end of the 1960s, Joseph Weber's announcements claiming detection of gravitational waves sparked the decisive entry of this group into the field with a three-branched approach: experimental detection, statistical analysis of the results and a deep theoretical footing in general relativity with the appointment of Jürgen Ehlers, coming from Pascual Jordan's school in Hamburg, one of the centers of the *renaissance* of general relativity. The Munich area group of Max Planck institutes provided the fertile ground for acquiring a leading position in the 1970s, facilitating the experimental transition from resonant bars towards laser interferometry and its innovation at increasingly large scales, eventually moving to a dedicated site in Hannover in the early 1990s.

The scales by then demanded the formation of international collaborations. An early pan-European initiative broke up into two major projects: the British-German GEO600, and the French-Italian Virgo. The Hannover group emphasized perfecting experimental systems at pilot scales, and never developed a full-sized detector, rather joining the LIGO Scientific Collaboration at the end of the century. In parallel, the Max Planck Institute for Gravitational Physics (Albert Einstein Institute) had been founded in Potsdam with a theoretical and computational emphasis. Both sites, in Hannover and Potsdam, became a unified entity in the early 2000s and were central contributors to the first detection of gravitational waves in 2015.

# Table of Contents

Acknowledgments

References

# 1. Introduction

In this contribution we explore the interplay between the *renaissance* of general relativity and the advent of relativistic astrophysics following the story of how gravitational wave detection found its place in the Max Planck Society, one of the world's most prestigious research organizations, based in Germany and conducting basic research in the natural, life, social sciences and the humanities. Based on this premise, we will then outline how the development of this research eventually led in 1995 to the building of the British-German interferometer GEO600, and to the foundation of the Max Planck Institute for Gravitational Physics, the Albert Einstein Institute.[4] Since the mid-1980s, the original aims of Max Planck researchers included to build a fully-sized 3-km gravitational wave interferometer, but in the early 1990s, they were forced to scale down to a pilot facility. Instead, GEO600 had a fundamental role in developing and testing advanced key technologies which contributed to the unprecedented sensitivity of the last generation interferometers Advanced LIGO and Virgo, which successfully detected for the first time gravitational waves in September 2015 (LIGO Scientific Collaboration & Virgo Collaboration 2016).

The Max Planck Society was founded in 1948 on the ashes of the Kaiser Wilhelm Society, whose origin dates back to the beginning of the past century. This unique scientific organization is formed today by more than 80 institutes and research facilities and it is principally financed by public funds. Over four decades, gravitational-wave research rose inside the Max Planck Society to prominence from its humble origins as a 'dark horse', backed by theoretical astrophysicists and particle physicists but generally marginalized by observational astronomers.[5] Our central focus is on the early decades up to the point when the field becomes firmly established and legitimate outside narrow circles, a period during which researchers in these fields circulate in small groups among different Max Planck Institutes initially in the Munich area (Physics and Astrophysics, Extraterrestrial Physics, Plasma Physics, Quantum Optics), before maturing into fully dedicated sites in Hannover and Potsdam, coinciding with the formation of the large international collaborations such as LIGO, Virgo and GEO600, and culminating in their unification under the Max Planck Institute for Gravitational Physics, the Albert Einstein Institute.

---

[4] The present contribution is resulting from our larger work on the history of astronomy, astrophysics and space sciences in the Max Planck Society performed within the Research Program "History of the Max Planck Society" (http://gmpg.mpiwg-berlin.mpg.de/en/, accessed 1/8/2019).

[5] For an account of the experimental search for gravitational waves, from the early days of the quest to the very recent past, up until 2004, when the book was first published, see (Collins 2004). The book focused in particular on the sociology of doing science and the growth from small-scale research to large-scale projects, but it also covered many aspects of the development and interpretation of gravitational-wave research in unprecedented detail. A book for a wider audience, reconstructing the history behind the first detection of gravitational waves on September 14, 2015, was written by Hartmut Grote, one of the scientists working at the German interferometer GEO600 (Grote 2018), its English version was published in 2019 (Grote 2019).

In the Max Planck Society, gravitational wave research began in the late 1960s, when general relativity was becoming "one of the most active and exciting branches of physics" based on the premises laid in the post-World War II period by the process dubbed "Renaissance of general relativity" (Will 1989, 7), which marked the return of Einstein's theory of gravitation to mainstream physics.

Recent scholarship has actually shown that the revival of the field started already in the 1950s as a consequence of a variety of elements, but it was mainly due to two factors, combining epistemological with sociological aspects: the discovery of the untapped potential of general relativity—as created by Einstein —as a tool for theoretical physics, and the emergence of a real community of relativists and cosmologists (Blum et al. 2015; 2016; 2017; Lalli 2017; Lalli et al. 2020; Lalli 2020).[6]

While this process was consolidating, towards the end of the 1950s, astronomical observations and calculations helped to spread the belief that general relativistic effects might be significant not only for cosmology, but might be of key importance also for interpreting the existence of violent events in the nuclei of galaxies and in isolated astronomical objects, as revealed by radio astronomy. The identification of celestial bodies of a new type, the quasi-stellar radio sources (*quasars)* announced in early 1963 (Hazard et al. 1963; Schmidt 1963; Oke 1963; Greenstein and Matthews 1963), the discovery of a rapidly pulsating radio source (*pulsar*) in 1967 (Hewish et al. 1968), as well as the serendipitous detection in 1964 of the Cosmic Background Radiation (Penzias and Wilson 1965)—a supposed relic of the very early universe—appeared to be astrophysical bodies or phenomena pertaining to the realm of Einstein's theory.

Quasars were soon connected either with collapsing "superstars" or with very compact remnants resulting from supernova explosions. And indeed, investigations about the final fate of super-dense stars had arisen since the 1930s the necessity of taking into account Einstein's general theory of relativity. In 1939, Robert Oppenheimer and Hartland Snyder focused the attention on the process of gravitational collapse itself (Oppenheimer and Snyder 1939; Bonolis 2017; Almeida 2020). For the first time, it became evident that this phenomenon is of basic importance for the understanding of the nature of space and time, being a unique process where a fully relativistic theory of gravitation could be seen at work and the validity of general relativity might be confronted with observational evidence. Oppenheimer and Snyder's pioneering and foundational work— in fact one of the cornerstones of relativistic astrophysics—was completely ignored at the time and almost forgotten during World War II. Only during the post- and Cold War period did implosion and explosion problems, related to the design of thermonuclear weapons, bring about renewed interest in investigations on highly dense stellar matter and on the abandoned problem of gravitational collapse within Einstein's theory. New tools, typical of post-war science, were now available: the impressive advances in nuclear science combined with the first powerful computers, designed to perform the complex calculations for thermonuclear weapons, were now used to calculate the equation of state of condensed stellar matter up to the endpoint of thermonuclear evolution.

---

[6] See also (Will 1993).

In the 1950s, two physicists—John Wheeler in the United States and later Yakov Borisovich Zeldovich in the Soviet Union—emerged from their respective hydrogen bomb efforts and took up where Oppenheimer and his collaborators had left off in 1939, at the outburst of the war. Wheeler rediscovered Oppenheimer's forgotten papers and was led to a systematic study of general relativity. As he recalled in his autobiography: "It was actually nuclear physics and quantum theory that drew me into relativity" and somewhere else he added "It is hardly possible for someone interested in nuclear physics and relativity, as I was and am, not to get interested in stars. My Princeton friend Martin Schwarzschild [the well-known astrophysicist, son of Karl Schwarzschild] drew me into stellar atmospheres. From there it was natural to fall, so to speak, into the center of stars" (Wheeler 1998, 228 and 292).[7] Wheeler even decided to teach a course to really delve into general relativity and thus, around his wide research project, Princeton became one the most lively research centers, contributing to fire up the great revival of general relativity. Zeldovich and Wheeler had remarkable parallel interests: Zeldovich, already one of the most influential USSR scientists, had a strong background in nuclear physics and had been one of the main creators of the first Soviet nuclear weapon. He thus saw that the physics of stars and the physics of nuclear explosions have much in common and reoriented his research agenda towards the physics of matter under extreme conditions and especially towards cosmological issues and like Wheeler had a strong influence on other theorists within his seminar (Sunyaev 2004).

---

[7] Martin Schwarzschild remembered how astrophysical problems were used to test computers also used for thermonuclear research. John von Neumann, who had built at Princeton the first electronic computer MANIAC, "was very interested to have a problem which was nonlinear and sufficiently complicated to really need the whole power of his machine, but where lots of hand computations for checks were available; and therefore the stellar evolution work, which I think von Neumann also considered interesting in itself, though not all that deeply — he thought that that was an excellent one. So actually next to the official major program, the meteorological dynamics for which the machine officially was funded, stellar evolution got the biggest share of time." Martin Schwarzschild: Interview by David DeVorkin and Spencer Weart, December 16, 1977, Session III. Transcript, Niels Bohr Library & Archives, American Institute of Physics, College Park, MD USA (from now on AIP),

These scientists, and other groups all over the world inspired a new generation of young theorists who laid the foundations for investigations on what is still one of the major challenges of relativistic astrophysics, namely: the equation-of-state at the center of a super-dense star, a main area of application of general relativity, together with gravitational waves and cosmology.

And indeed Wheeler further recalled: "What interested me was not the center of an ordinary star like our Sun, cooking away and generating thermonuclear energy. I was interested in the center of a cold dead star" (Wheeler 1998, 292). In the second half of the 1950s, assembling all the available theoretical information, Wheeler and his collaborators Ken Harrison and Masami Wakano constructed a semi-empirical equation of state for matter in its absolute ground state to the endpoint of thermonuclear evolution, at all stages of compression, up to supranuclear densities. Computer simulations confirmed previous theoretical results obtained during the 1930s by Subrahmanyan Chandrasekhar and by Oppenheimer and collaborators, that for the first time were brought together into a single overall picture (Wheeler et al. 1958, Fig. 12). Wheeler's group presented this work at the eleventh Solvay Conference held in June 1958 devoted to the "Structure and evolution of the Universe", that gathered some of the most distinguished astronomers and experts in cosmology and general relativity.[8] The topic chosen

---

https://www.aip.org/history-programs/niels-bohr-library/oral-histories/4870-3, accessed 30/7/2019. Schwarzschild also remembered that Wheeler was joined for a whole year by the theoretical astronomer Louis Henyey, who spent the period 1951-1952 at Princeton University where he was involved in the classified defense work on Project Matterhorn, the US top secret project to control thermonuclear reactions. Martin Schwarzschild: Interview by William Aspray, Princeton, November 18, 1986. Transcript. N. J. Charles Babbage Institute. Retrieved from the University of Minnesota Digital Conservancy, http://hdl.handle.net/11299/107629, accessed 7/7/2019. Henyey realized that what he had learned at Princeton from von Neumann was extremely useful for the stellar interior and developed what came to be known as the 'Henyey method', which became the standard tool for the theory of stellar interior (Henyey et al. 1959).

[8] See especially the session "Matter-energy at high density; end point of thermonuclear evolution" in their

for the Solvay congress meant an acknowledgment of cosmology as part of physics, thus mirroring the growing influence of results coming from radio astronomy and from nuclear astrophysics on the debates on cosmological models.[9] It was the first of this kind within the Solvay series, but it was also the first in establishing an official connection between astrophysics and general relativity. More in general, at the end of the 1950s, the considerable revival of interest in compact stars, and in the properties of matter at high densities, led to discussions and investigations on topics such as neutron stars or the possibility of gravitational collapse to a singularity in space-time.[10]

Technological progress during World War II had opened new horizons in the study of astronomy and the realm of radio stars and very distant radio galaxies had become a subject of investigation. During the 1960s, the discovery of quasars—or "superstars" as they were originally called—but in particular of the pulsars—discovered in 1967 and immediately identified as neutron stars—for the first time appeared to offer the chance to solve the conflict

---

contribution to the proceedings: "In seeking the consequences of Einstein's theory for the structure and evolution of the universe we have been forced to consider what happens during contraction. Such implosion can be expected to lead not only to the dynamic instabilities just discussed, but also to unprecedentedly high densities of matter and radiation. Such densities pose unsolved problems to general relativity and elementary particle physics" (Wheeler et al. 1958, 124). In his relevant report at the Solvay conference, Ambartsumyan suggested that the nuclei of galaxies are often centers of large scale activity proceeding in different form and that the radio galaxies are not the products of collision of galaxies, as was accepted at that time, but are systems in which ejections from the nuclei of tremendous scale take place: "Apparently we must reject the idea that the nucleus of a galaxy is composed of common stars alone. We must admit that highly massive bodies are members of the nucleus which are capable not only of splitting into parts that move away at a great velocity but also of ejecting condensations of matter containing a mass many times exceeding that of the Sun […]" (Ambartsumyan 1958, 266).

[9] The opening lecture was given by Georges Lemaître, giving a full account of his theory of the "primaeval atom". George Gamow, instead, was not invited. He was the main supporter of a universe emerging from a singularity, mockingly termed "big bang" by Fred Hoyle, the main opponent of this theory, who gave first a talk on the steady state theory and then a second one on the origin of elements in stars. Radio astronomical observations as a source of information on the structure of the Universe were also discussed during the Solvay conference.

[10] The possibility of using the new computing facilities to investigate the physics of supernova explosions and of the formation of the elements, actually led Alastair G. Cameron, working at Chalk River, Canada's largest nuclear science and technology laboratory, to construct 20 neutron star models by integrating the general relativistic equations of hydrostatic equilibrium of the neutron gas, which discussed the transformation of neutrons into other baryons like hyperons at very high densities and showed, among other results, that neutron stars could be probable products of the supernova process (Cameron 1959). Towards the end of the 1950s/early 1960s, discussions on the question of the equation of state of matter at ultrahigh densities in connection with the problem of the last stage of the evolution of heavy stars was intensifying. See for example (Salpeter 1960; Zeldovich 1962a; Zeldovich 1962b) and in particular David Beckedorff's dissertation on the terminal configurations of stellar evolution made under the supervision of Charles W. Misner (Beckedorff 1962). Apart from Cameron's article which attracted attention of theoreticians towards neutron stars, other pioneering attempts had been made by taking into account the presence in dense stars of various elementary particles also through investigations on the nuclear interaction between nucleons and hyperons, some of which had been discovered in cosmic rays (Ambartsumyan and Saakyan 1960; Hamada and Salpeter 1961).

between the already developed theory of the structure, evolution and final fate of massive compact stars and the fact that such objects, since a long time discussed as theoretical entities by a few physicists, had not yet been observed. As opposed to cosmology, astrophysics had remained "non-relativistic" up to the early 1960s. The processes during which massive cosmic bodies and systems of bodies reach a relativistic stage became a central subject of the emerging field for which the term "relativistic astrophysics" was coined in 1963 (Robinson et al. 1965). This brand new field was creating—and in turn was being created by—the growing dialogue and co-operation between astrophysicists and relativists on the widening scenario set up by the *renaissance* process already dawning into the "astrophysical turn" of general relativity (Blum et al. 2018; see also Lalli et al. 2020 and Chapter 1 in Blum et al. 2020).[11]

Relativistic astrophysics, however, could start in earnest only after the discovery of the pulsars. As a new class of radio sources characterized by the emission of short pulses of radiation having an extremely constant repetition frequency, immediately identified as rotating magnetized neutron stars (Gold 1968),[12] such super-dense stellar cores left behind after a supernova explosion, dramatically brought back to the attention of a large section of theoretical and experimental astrophysicists the basic issues of the physics of gravitationally collapsed objects.[13] The clear evidence for the existence of such compact astrophysical objects, for which it was recognized that Einstein's theory of general relativity plays an essential role, made even plausible the reality of more exotic theoretical entities such as "black holes", as stressed by Wheeler and Ruffini: "No one who accepts relativity has seen any way to escape their existence" (Ruffini and Wheeler 1971, 39).[14]

---

[11] The process of the renaissance of general relativity has been analyzed in the already mentioned publications (Blum et al. 2015; 2016; 2017). Kip Thorne, on the other hand, designated the period 1964-1975 as the "golden age" of general relativity (Thorne 1994, 258-299), which he actually identified with the explosion of interest towards black hole research. For a different point of view on the concepts of "low water mark between 1925 and 1955", "renaissance" and "golden age" of general relativity, see (Goenner 2017b).

[12] Pulsars are characterized primarily by the emission of sharp flashes of radio waves at almost exactly maintained time intervals. Towards the end of the same year, two pulsars were found in known sites of supernovae explosions, the source PSR 0833 in the Vela Remnant and the source NP 0532 in the Crab Nebula. This led to the general acceptance of the rotating neutron star hypothesis for the basic nature of these objects, as suggested by Thomas Gold.

[13] The authors pointed out that the radiation might be "associated with oscillations of white dwarf or neutron stars," thus focusing attention on a very dense astrophysical object. For a more general treatment see (Hewish 1970).

[14] The term "black hole" began to circulate and was officially launched by John Wheeler in 1968 (Wheeler 1968). However, it is not clear who used it first, although it appears that in reality it circulated as early as September

As remnants of gravitational collapse following a supernova explosion—a main candidate event for the emission of gravitational waves—pulsars were immediately recognized as promising sources of such waves, whose search was going to become a main experimental branch of relativistic astrophysics as well as a fundamental test of Einstein's theory outside the solar system.

In the wake of the burst of interest brought by these major multifaceted developments, and thanks to the visionary support of its director, the well-known astrophysicist Ludwig Biermann, the Max Planck Institute for Astrophysics[15] in Munich was among the first to enter experimental gravitational wave research, taking advantage of the existing technical and experimental capabilities headed by the pioneer of electronic computing machines Heinz Billing. In parallel, general relativity was established as a new research field at the institute, appointing a series of renowned experts. Through their strong participation in the early debates on gravitational wave detection by Joseph Weber, Munich researchers acquired a dominant position at an early stage, supported by the recognition for their theoretical and experimental work. This privileged position was then mobilized for the timely reconversion of their experimental program from resonant bars to laser interferometry. In this venture, they took significant advantage of the shared resources, of expertise and infrastructure from other nearby Max Planck Institutes in the Munich area. In fact, between the early 1970s and the late 1990s, the gravitational waves research group was part of different institutes of the Munich family, before being finally absorbed in 2002 by the Max Planck Institute for Gravitational Physics—the "Albert Einstein Institute"—already established in Potsdam since 1995.

This contribution tells the story of the origins and itinerant life of this research group up until it found a home at a new dedicated Max Planck Institute. And while it gives an account of how gravitational physics entered the Max Planck Society, and sufficient allusions are made to the large-scale global collaborations that over the past two decades eventually resulted in the

---

1963, during the first Texas conference, as reported in the issue dated January 24, 1964 of *Life* magazine by Al Rosenfeld, *Life*'s science editor, who had heard the term mentioned during the symposium (Siegfried 2013).

[15] Historically, Ludwig Biermann's Institute for Astrophysics was part of a larger entity called the Max Planck Institute for Physics and Astrophysics headed by Werner Heisenberg. These institutes shared the same building and benefitted from being part of a common umbrella "Max Planck Institute" which also included since 1963 the Institute for Extraterrestrial Physics. Only in the early 1990s these were formally divided into separate, fully-fledged Max Planck Institutes. In this article however, to distinguish it clearly from other organizations around the world, we use the term "Max Planck Institute for Astrophysics" to refer to the entity headed by Biermann and his successors.

successful detection of gravitational waves, this contribution focuses mainly on the early phase of the story. Further detail on the 21st century stage of this global dynamic of competition and collaboration constitute ongoing research by many historians and sociologists since the successful detection of gravitational waves in 2015. Those studies can further elucidate the dynamic by which Germans contributed to instrumental developments while renouncing the infrastructural protagonism to the American (LIGO) and French-Italian (Virgo) collaborations.

**2. Relativistic astrophysics, quasars, and the beginning of the 'golden age' of general relativity**

The lead up to gravitational wave experiments undertaken in the early 1970s at the Max Planck Institute for Astrophysics was mainly rooted in the constantly growing interest within Ludwig Biermann's group in the new field of relativistic astrophysics which radically transformed our view of the universe from the early 1960s onward. During the previous decade, the advent of radio astronomy had revealed that much in the universe is of an explosive nature and that violent events exist in galactic nuclei (Burbidge 1956; Mayer et al. 1957).[16] Astrophysicists had tried to understand the source of the tremendous energy stored in cosmic rays and the magnetic fields of some powerful radio galaxies (Burbidge 1959).

From the late 1940s/early 1950s, the problem of the origin of high-energy cosmic ray particles and the mechanisms accelerating them with the related emission of synchrotron radiation, the source of the radio signals, had been a topic of great interest for Biermann. The inner workings of the Sun and stars had been his specialized area since the beginning of his scientific career during the 1930s and, as a result, their structure and evolution had constantly been one of the key research topics at the Max Planck Institute for Astrophysics. In particular, compact stars had attracted his attention at least since 1931, when he participated in the hot debate about theories on super-dense matter in white dwarfs, which became a subject of correspondence between Biermann and Pascual Jordan in the 1940s (Biermann 1931).[17] As can be inferred

---

[16] The idea that synchrotron radiation could supply an efficient mechanism by which individual sources (both galactic and extragalactic) could radiate large radio powers had a great effect also in optical astronomy. The discovery of linearly polarized light emission from the Crab Nebula—confirming the synchrotron nature of radio emission—meant that astronomers had a new tool for studying high-energy processes, thus becoming more oriented towards a high-energy universe manifested in the various radio sources with optical identification.

[17] On these developments see (Bonolis 2017). Biermann's correspondence with Pascual Jordan in 1946 was also about internal constitution of dense stars with neutron cores, a result of Jordan's extension of his interest in cosmology to astrophysics. On October 2, 1946, Jordan was asking Biermann's opinion on these issues (Archives for the History of the Max Planck Society [from now on AMPG], III. Abt., ZA 1, Nachlass Ludwig Biermann [in

both from publications and correspondence, during the 1940s and 1950s Biermann worked on topics involving astrophysical plasmas and magnetic fields in space—in general on cosmical electrodynamics—with his collaborators, notably Reimar Lüst, who would later become Director of the Max Planck Institute for Extraterrestrial Physics, and Arnulf Schlüter, who would lead the Institute for Plasma Physics (Biermann and Schlüter 1950; Biermann and Schlüter 1951).[18] Growing evidence for the existence of relativistic plasma as an essential, major component of the universe—which had been a popular research topic at the Institute for Astrophysics—triggered the explosive growth of high energy and relativistic astrophysics since then. It became also clear that radio galaxies were among the most distant objects in the universe, while detection of polarized radiation from the Crab Nebula at optical wavelengths confirmed that radio emission was due to the synchrotron mechanism. The jets of the galaxy Virgo A, too, turned out to be synchrotron radiation emitted by ultra-relativistic electrons spiraling in magnetic fields. During a meeting of radio astronomers in Paris which took place in summer 1958, immediately after the Solvay conference, the astrophysicist Geoffrey Burbidge discussed the implications of synchrotron radiation coming from Cygnus A and showed that the energy needed to produce the high-energy particles was much greater than the expected energy from a collision of galaxies as hypothesized by the astronomer Walter Baade. It appeared that the nuclei of galaxies might host the necessary source of energy for such powerful processes. Such theoretical insight led Burbidge to stress that enormous energies were at stake (Burbidge 1959).

The realization that the energy released within strong radio sources can exceed an energy equivalent of millions of solar masses soon led William Fowler and Fred Hoyle to explore the

---

the following, Biermann's papers will be cited in the abbreviated form NLB], No. 2). At that time, Jordan wrote a book on the constitution of stars, summarizing work published during the war, dealing in particular with super-dense matter. The book was reviewed by Biermann (Biermann and Jordan 1947). Biermann himself was working on super-dense white dwarfs (Biermann 1948). Jordan's interest was then turning to general relativity, creating the premise for future developments of this discipline in West Germany.

[18] Biermann was considered to be an expert, and as such was asked to write a review article on this subject (Biermann 1953). Still in 1965, Biermann wrote to the Soviet astrophysicist Vitaly Ginzburg, who had been one of the first to theorize about the phenomenon of synchrotron emission and who continued to be focused on the problem of the origin of cosmic rays: "The publication of several new papers which seem relevant to questions of the origin of cosmic rays reminds me that I have still omitted to thank you for the copy of your and Dr. Syrovatsky's valuable monograph on the origin of cosmic rays, which was sent to me by the publisher. This field of research is in such an active development that it is most useful to have such a fine description of the present state of our knowledge. It will be most interesting to see how the growing body of information on the quasi-stellar sources on the one hand, and on more extended sources such as IC 443 on the other hand, will affect our ideas on cosmic rays in the years to come." L. Biermann to V. Ginzburg, April 8, 1965, AMPG, NLB, No. 31.

possibility that "at the centers of the galaxies there are *star-like objects* with masses ranging from about $10^5$ up to about $10^8$ solar masses for abnormal galaxies [added emphasis]." Fowler and Hoyle's opinion was that "only through the contraction of a mass of $10^7 - 10^8$ solar masses to the relativity limit can the energies of the strongest sources be obtained" (Fowler and Hoyle 1962, 170). This article appeared in August 1962, but in the meantime, Hoyle and Fowler took a further step. In February 1963 they argued that nuclear energy could not be the key to the problem, being unable to maintain sufficient internal pressure even to provide support against gravity for such massive astrophysical objects and observed that gravitational energy, instead, could be of decisive importance for bodies in that range of masses. The energies demanded by the strong sources were "so enormous as to make it clear that the relativity limit must be involved." As this limit was approached "*general relativity must be used*" [added emphasis] (Fowler and Hoyle 1963, 535). "The conclusion was now clear; that at a certain stage of its contraction (at about the size of the whole solar system) a very massive object would *implode* catastrophically, in about 100 seconds" (Hoyle 1963, 682).

Soon after, in the following March, Fowler and Hoyle's suggestions appeared to materialize in the "star-like" objects—with a very large redshift and corresponding unprecedented large radio and optical luminosities—whose identification was announced in four consecutive articles in *Nature* (Hazard et al. 1963; Schmidt 1963; Oke 1963; Greenstein and Matthews 1963). The dramatic recognition of these unusual objects was the result of a fruitful collaboration between radio and optical astronomers. The former provided precise positions of radio sources, which were then identified with star-like objects on photographic plates. In recognition of their small size, they were called *quasi-stellar* radio sources, soon renamed *quasars*.

The most pressing problem in astrophysics at the time became how to explain the mechanism whereby the most bizarre and puzzling objects ever observed through a telescope to date, which proved to be the most powerful energy sources in the sky, managed to radiate away the energy equivalent of five hundred thousand suns in short order. It was immediately connected with what Fowler and Hoyle had suggested in their February article, only a few weeks before the announcement: *gravitational collapse might be the driving force behind the large amount of energy emitted by strong radio sources*. Since such enormous energies must be emitted by regions less than one light-week across, collapsed objects became candidates for the engine of quasi-stellar radio sources. Their unusually high redshifts showed that they were the farthest objects detected in the universe. The most direct explanation for such large redshifts was that the quasars were extragalactic, their significant redshift reflecting the Hubble-Lemaître expansion of the universe. Fowler and Hoyle's proposed mechanism involving gravitational collapse—a purely relativistic phenomenon at the time not yet fully understood—turned a spotlight on the bonds between general relativity, astronomy, and astrophysics.

In December 1963, the first international Texas Symposium on Relativistic Astrophysics was held in Dallas, organized by three relativists: Ivor Robinson, Alfred Schild, and Engelbert Schücking. This event took place at a time when the complex process developing since the aftermath of World War II, which had put in motion the "renaissance" of Einstein's theory after a long period of stagnation, was being completed. After remaining cut off from mainstream physics for a generation, this formerly dispersed field was attracting an increasing number of practitioners, becoming the basis for the standard theory of gravitation and cosmology. New connections were now on the verge of being established with astrophysics and physical cosmology, through which general relativity would enter its "astrophysical turn," becoming an established branch of physics.

The first Texas event stemmed from the idea of having a small conference as an occasion to "put on the map" the recently created Southwest Center for Advanced Studies in Dallas,[19] which was part of a larger project aiming to promote general relativity as a well-established research field at the University of Texas in Austin and in Dallas itself. The renowned relativist Ivor Robinson had just been appointed head of the Mathematical Physics Division of the recently created Southwest Center for Advanced Studies in Dallas, a successful result of Alfred Schild's far-sighted initiative—as leader of the Center for Research in Relativity Theory at the University of Texas in Austin (Lalli 2017).[20] In 1962, Schild had also got Engelbert Schücking an associate professorship in the Austin mathematics department and, as Schücking himself recalled, "in the summer of 1962, while attending Andrzej Trautman's relativity conference in

---

[19] See XXVII Texas Symposium December 8-13, 2013, Dallas, *Roundtable Discussion – Wednesday, December 11th "Recollections of the Relativistic Astrophysics Revolution"* 2013, https://nsm.utdallas.edu/texas2013/events/, accessed 31/7/2019. This special event dedicated to the 50th anniversary of the first Texas Symposium on Relativistic Astrophysics was organized during the XXVII Texas Symposium held in Dallas, December 8-13, 2013. It gathered veteran scientists recalling the circumstances that led to the first Texas meeting in 1963 and reflecting on the subsequent impact of such conferences. Schücking was not able to participate, but he shared his own recollections in a video (Schücking 2013).

[20] Schild had studied physics with Leopold Infeld, one of Einstein's disciples and collaborators, writing a thesis on cosmology.

Warsaw, Poland […] we persuaded Roger Penrose, Roy Kerr, Ray Sachs, Jürgen Ehlers, Luis Bel and others to flock to the newly created center of gravity in Austin" (Schücking 1989, 46–47).

Both Schücking and Ehlers had studied general relativity with Pascual Jordan, one of the pioneers of quantum physics, who had formed a research group in Hamburg back in the early 1950s, which was one of the seeds fertilizing the renaissance of general relativity.[21] Ehlers himself, emphasizing Jordan's wide-ranging interests, later recalled:

> […] it was not astonishing that Jordan's seminar on General Relativity and cosmology which began in the mid-fifties and was carried on at the university of Hamburg for about fifteen years attracted many talented students, several of whom later attained professorships or research positions in physics or mathematics. At the post war time when General Relativity had almost been forgotten not only in Germany, Jordan recognized the importance of this field for future research. Due to Jordan and his students and collaborators the renaissance of General Relativity around 1955 took place not only in Syracuse, Princeton, Paris, London, Dublin, Leningrad and Warsaw, but also in Hamburg. Without this germ-cell of General Relativity in Germany a relativity group would presumably not have been created at a Max Planck Institute […].[22]

As members of Jordan's group, Ehlers and Schücking became themselves key actors in the establishment and further evolution of the renaissance of general relativity. As we will see later, Ehlers' choice to leave Austin and go back to Europe is instrumental in our story.

On a hot July afternoon in 1963, Robinson, Schild, and Schücking were celebrating their reunion drinking strong martinis around a swimming pool in Dallas (Schücking 1989).[23] The idea of a meeting which could attract the attention on the new Southwest Center for Advanced

---

[21] Ehlers earned his PhD in Physics at the University of Hamburg in 1958 with a dissertation entitled "Konstruktionen und Charakterisierungen von Lösungen der Einsteinschen Gravitationsfeldgleichungen" (Constructions and characterizations of solutions to Einstein's gravitational field equations). He obtained his habilitation in the same university in 1961 and for a short time held a position as assistant professor at the Christian Albrechts University in Kiel before moving to Syracuse University, New York, as a research associate that same year. After working several years with Alfred Schild's group, in 1965 he became associate professor at Austin, Texas, and since 1967 he was full professor there until 1971, when he moved back to Germany.

[22] Jürgen Ehlers, "Pascual Jordan – Originator of Quantum Field Theory and Founder of a Relativity School" in (Schutz 2003, 13). See also (Goenner 2017a; Lalli 2017). For Ehlers' obituary, see (Allen et al. 2008).

[23] At the beginning of his recollections, Schücking also mentioned a party to which all of them participated in that same evening and which took place at Manfred Trümper's. The latter had obtained his PhD in Hamburg with Pascual Jordan, and after being Bergmann's postdoc and subsequently becoming associate professor at the University of Texas, he would later become a member of Ehlers' research group on general relativity at the Max Planck Institute for Astrophysics.

Studies emerged and crucially, Engelbert Schücking suggested organizing a conference on the "mysterious star-like objects" which had recently come to the fore, and which were supposed to be connected to general relativity.[24]

They immediately involved Peter Bergmann in the organization of the conference. Bergmann, an influential relativist who had been Einstein's research assistant at the Institute for Advanced Study in Princeton since 1936, and had joined Syracuse University in 1947 also being active with a research appointment at the Yeshiva University, New York. Since the 1940s, he had established a center for relativity research at Syracuse University which attracted all the leading relativists and became one of the very first active research group on general relativity in the post-World War II period.

Consequently, the International Symposium on Gravitational Collapse and Other Topics in Relativistic Astrophysics, ultimately a "monster conference" hosting about 300 scientists, the first of a long series of Texas Symposia, was held in Dallas from December 16 to 18, 1963 (shortly after John Kennedy's assassination). Bringing together optical and radio astronomers, theoretical astrophysicists, and general relativists, it marked the start of a new era bridging the gap between the still exotic world of general relativity and the realm of astrophysics. Moreover, it officially opened the discussion on topics ranging from neutron stars to the possibility of gravitational collapse or a singularity in space-time, setting the stage for a dialogue between different scientific communities (Robinson et al. 1965).

---

[24] L. Marshall: Interview by Alan Mitchell, June 4, 1978. Transcript. Graduate Research Center of the Southwest, Center for Advanced Studies (SCAS) Collection, Special Collections and Archives Division, The University Archives, University of Texas at Dallas, Box 3, Folder 23. Schücking had been Pascual Jordan's student in Hamburg from 1952 onward. Astronomy had been one of his passions since he was a child and, at the age of 14, he was counting sunspots for Zurich Observatory. His first appointment was to Hamburg Observatory, at Bergedorf, where Walter Baade had worked since 1919, before emigrating to the U.S. in 1931, and where Biermann himself had spent some time from the end of the war up to 1947, when he moved to Göttingen at Heisenberg's Max Planck Institute for Physics. At the time Bergedorf was considered "perhaps the principal observatory of Germany" (Kuiper 1946, 267). Schücking worked there from 1941 to 1962 with the astronomer Otto Heckmann, its director and head of the department of astronomy at Hamburg University, also known for his studies of relativity and cosmology. Heckmann and Schücking later co-authored several influential articles on relativistic and Newtonian cosmology. See p. viii and Schücking's Curriculum Vitae in (Harvey 1999, 515) and see also (Goldberg and Trautman 2018). Schücking recalled how his "interest in the physical aspects of cosmology was sparked" in 1955, when Walter Baade—perhaps the greatest observational astronomer of his time—had come from Pasadena to celebrate the inauguration of the Schmidt telescope. Baade had worked at Hamburg Observatory at Bergedorf from 1919 to 1931 and was then at Mount Wilson Observatory, where, together with Rudolph Minkowski, he had identified the optical counterparts of various radio sources, including Cygnus A, one of the brightest sources in the sky. The nature of such a source and the origin of its intense power was at the center of a large debate between astronomers, radio astronomers, and astrophysicists during the 1950s. Schücking was "absolutely fascinated by Baade's research" as he remembered in the mentioned video (Schücking 2013). He continued to follow developments in radio astronomy very closely during the 1950s, participating in conferences and meetings where radio sources and cosmological implications were continuously debated. In particular, along with Heckmann, he participated in the Solvay Conference of 1958 as well as in the Paris Symposium on Radio Astronomy of the International Astronomical Union held in July-August of that year. In 1961, Schücking visited the U.S., spending most of his time in Syracuse, with Bergmann, but also having the opportunity to meet Felix Pirani and Fred Hoyle during his first visit to England. He then attended several conferences, where he met Georges Lemaître, Margaret and Geoffrey Burbidge, and all the most influential US astronomers and astrophysicists. In 1962, he was appointed Associate Professor at the University of Texas at Austin. This background clarifies why Schücking was so quick to grasp the importance of the new discovery of quasars and the related astrophysical and cosmological implications.

A general consensus began to emerge from the awareness that general relativistic effects can play a dominant role in astrophysics and that astrophysical objects exist in the universe that are understandable *only* in terms of Einstein's theory. The classical works on gravitational collapse by the Indian physicist B. Datt[25] and especially that by Robert Oppenheimer and Snyder published in the late 1930s (Datt 1938; Oppenheimer and Snyder 1939) and almost forgotten were being rediscovered and discussed, with acceptance of the situation that "stars with masses greater than the critical mass can reach a stage of catastrophic implosion in which general relativity becomes dominant" (Hoyle et al. 1964, 910).[26]

---

[25] There is very scant information on Datt. The well-known Indian astrophysicist Jayant V. Narlikar, included a specific section on Datt, entitled "Who was B. Datt?", in his article on the early days of general relativity. According to his enquiries, Datt belonged to the famous Presidency College and was a favourite student of the theoretician N. R. Sen, who had done his PhD in Berlin with Max von Laue, and later established the first general relativity school at the University of Calcutta, and worked on solutions of Einstein's equations with mathematically significant properties. In this regard, see Sen's 1934 article on the stability of cosmological models republished in the journal *General Relativity and Gravitation* (Sen 1997). As a post-doc working on gravitational collapse, Narlikar found "Datt's approach quite general" and in fact he discovered that the famous Landau-Lifshitz text book *Classical Theory of Fields* (2nd edn) had a reference to Datt's paper whose method was actually followed in the book. Narlikar compared Oppenheimer and Snyder's paper to Datt's contribution: "The Oppenheimer-Snyder paper is generally regarded as the pioneering work on spherical massive objects contracting with increasing inward speed. Datt, however, kept his approach general; thus giving solutions not only of contraction but other motions too. More importantly, he had seen the significance of comoving coordinates in solving such problems." Interestingly, his paper was completed in September 1937, more than one year before the appearance of the Oppenheimer-Snyder paper. As to why no later work by him can be found, Narlikar reported what he had learned from Somak Raychaudhury: it was due to Datt's untimely death in the course of a surgery in 1940 (Narlikar 2015, 2216).

[26] Such papers ("the classical implosion problem") were the starting point of a discussion by Hoyle, Fowler and the Burbidges (Geoffrey, a theoretical astrophysicist and his wife Margaret, an astronomer) on various situations "envisaged as the final stages of evolution of a star which reaches the end point of thermonuclear evolution with a mass greater than the mass which can be supported either by degenerate electron pressure or degenerate neutron pressure" (Hoyle et al. 1964, 910). The process of gravitational collapse had been studied in detail since the 1950s by John Archibald Wheeler and his colleagues and presented at the first Texas conference. It was published as a separate volume from the general proceedings (Harrison et al. 1965).

## 3. A privileged role for Ludwig Biermann's Institute for Astrophysics through theory

Rudolf Kippenhahn, who would later become Biermann's successor as Director of the Max Planck Institute for Astrophysics, was heavily involved in research into the structure and evolution of stars, particularly with computer simulations.[27] He had participated in the first Texas conference, which officially launched the brand-new field of relativistic astrophysics, merging two seemingly distant fields, so far removed that the organizers had to invent a new label for this.

In 1964, the detection of the cosmic microwave background radiation by Arno Penzias and Robert Woodrow Wilson (Penzias and Wilson 1965), together with the interpretation by Robert Dicke and his associates of such radiation as a signature of the Big Bang, provided a new element in favor of the Big Bang cosmological theory and marked the start of a new era for physical cosmology.[28]

In lessons held in 1965 at the Enrico Fermi summer school in Varenna, Italy—two years after the first Texas symposium—Kip S. Thorne clearly outlined how the early developments of relativistic astrophysics were connected with the ongoing "astrophysical turn" of general relativity:

> Astrophysics and general relativity influenced each other very little during the long period between the first few years of relativity theory and about 1963. In fact, during that period the absence of any extensive experimental or observational phenomena in which general relativistic effects might be important tended to insulate Einstein's theory from all other branches of physics. However, during the last three years a marked change has begun to occur: The discovery and investigation of quasi-stellar radio sources, of explosions in galactic nuclei, and of X-ray emission from supernova remnants have suggested to astrophysicists that strong gravitational fields might, after all, play an important role in astrophysical phenomena. At the same time, major advances in the techniques of radio and optical astronomy have enabled astronomers to begin to determine the cosmological structure of the universe—which structure is believed to be governed by general relativity— and the development of powerful new experimental techniques has made possible new and improved tests of Einstein's theory. Because of these developments, strong gravitation physics as described by general relativity is rapidly becoming of interest to astrophysicists, and astrophysics is rapidly becoming of interest to relativists (Thorne 1966).

---

[27] In the early 1960s, Kippenhahn used for such pioneering simulations a new version of the Henyey code he had learned in the U.S. and which had been improved at the Institute for Physics, running it on the computers built there by Heinz Billing during the 1950s, when no commercial electronic computers were still available.

[28] The phenomenon had already been predicted at the end of the 1940s by George Gamow's collaborators Ralph Alpher and Roger Herman within investigations on the origin of chemical elements in the Universe (Alpher and Herman 1948). The subsequent interpretation of the cosmic microwave background radiation (CMB) as the redshifted cool remnant of the thermal radiation of the hot early phases of the Big Bang, whose presence was to be expected if the expansion of the universe could be traced back to a time when the temperature was of the order of $10^{10}$ K, opened one of the most fruitful areas in observational cosmology (Dicke et al. 1965). CMB provides an omnipresent radiation background dominating all-sky images at millimeter and sub-millimeter wavebands, which are also characteristic of a wealth of molecular lines such as those observed in regions of star formation. On Robert Dicke's pioneering activity in the experimental study of gravity, see (Peebles 2017). If the radiation were truly the relic radiation from the early hot universe in thermal equilibrium, then it would have the famous blackbody spectrum whose formulation by Max Planck had initiated quantum theory in 1900. Measurements made by the COBE satellite, very precisely fitting with the expected blackbody curve with T=2.726 K as predicted by the hot Big Bang theory, definitely corroborated the blackbody nature of the CMB spectrum, firmly establishing in the early 1990s that CMB is the relic thermal radiation from the primeval fireball that began our observable universe about 13.7 years ago (Mather et al. 1994). John C. Mather and George F. Smoot were awarded the Nobel Prize in Physics 2006 "for their discovery of the blackbody form and anisotropy of the cosmic microwave background radiation." The Nobel Prize in Physics 2006. NobelPrize.org. Nobel Media AB 2020. https://www.nobelprize.org/prizes/physics/2006/summary/, accessed 29/2/2020.

Thorne, who was a member of the group led by John Wheeler at Princeton, one of the major centers of general relativity in the 1950s-1960s,[29] fully lived the emergence and development of relativistic astrophysics during the 1960s-1970s, with a strong focus on relativistic stars, black holes and gravitational waves.

This backdrop—connecting a new phase in the study of general relativity and gravitation to the emergence of relativistic astrophysics—and the new perspectives opening up in general for astrophysical sciences, are definitely mirrored by research activities performed at Ludwig Biermann's Institute for Astrophysics. The section "Aufbau und Entwicklung der Sterne" (Stellar structure and evolution) suddenly became the most extended in the 1964 research report where, for the first time, a few lines dedicated to general relativity appear (Biermann and Lüst 1965).

From 1964, the young researcher Peter Kafka began to work on topics related to general relativity and cosmological questions at Biermann's institute in Munich (Biermann and Lüst 1965, 61).[30] He investigated the problem of gravitational collapse in general relativity, but in

---

[29] See Blum and Brill's chapter in the forthcoming volume *The Renaissance of General Relativity in Context* (Blum et al. 2020).

[30] Peter Kafka later recalled that at the time he made his *Diplom* in Physics, Arnulf Schlüter, one of Biermann's older collaborators, who had become Director of the Max Planck Institute for Plasma Physics in 1965, had developed an interest in general relativity and asked him to work on this topic for his dissertation. Peter Kafka: interview by Michael Langer, March 27, 1999. Live-Gespräch-Sendung "Zwischentöne," Deutschlandfunk, http://www.gegen-den-untergang.de/zwischentoene1999.html, accessed 11/4/2018. At that time quasars were discovered and so it became quite clear that general relativity would have a growing role in astrophysics and in cosmology and as an expert in such topics Kafka got a stable position at the Max Planck Institute for Astrophysics

particular he explored the space-time distribution of the quasars and radio galaxies as deduced from observational evidence (Biermann and Lüst 1966, 71). From radio astronomical observations it appeared that there were relatively more quasars at larger distances, and so that must mean they were more common in the early life of the universe. This could be explained as an effect of its evolution: if their redshifts were of cosmological origin, quasars—whose very nature was still object of debate—must have existed only very far away in time and space, contradicting the perfect cosmological principle, which was at the core of steady-state cosmology. The counting of radio quasars as recognized sources at cosmological distances might thus help to confirm the existence of the Big Bang model, ruling out the steady state model of the universe, according to which the expanding universe would maintain a constant average density, with matter being continuously created to form new stars and galaxies.[31] However, as Kafka pointed out in his article in *Nature*, there was disagreement "about the meaning of relations between the observed numbers, redshifts and brightnesses of quasars." He then concluded: "no decision can be made, *from a statistical count of quasars*, between steady state and other cosmological models [added emphasis]" (Kafka 1967).[32] Through Kafka,

---

(Biermann and Lüst 1966). At the institute, cosmological questions related to the distribution of clusters of galaxies were also examined by G. O. Abell of the University of California and associated with the Mount Wilson and Palomar Observatories, an expert in extragalactic studies, then guest of the Institute. On such research issues see also (Kafka 1968a).

[31] The static universe proposed by Fred Hoyle and, independently, by Hermann Bondi and Thomas Gold, rejected the idea of an initial singularity, maintaining that a steady-state universe could be compatible with the drifting apart of galaxies if new matter (approximately one hydrogen atom per cubic kilometer per year) were continuously generated in the intergalactic space. Since the mid-1950s, complete new catalogues of radio sources had shown that the number of intense sources increased with distance, while from the steady-state theory they were expected to be uniformly distributed throughout the universe. Apparently the most distant objects of the universe, quasars, had an impact in cosmology. If the high redshift of observed quasars was of cosmological origin, it meant that they were at distances such that the universe was much younger than it is now when the radio waves were emitted. This implied that galaxies produced more radio waves in the past, and thus began to call attention to the conflict between the Big Bang as a theory of cosmic expansion from a hot early universe and the steady-state cosmology, according to which the observable universe is basically the same on the large scale at any given time, a view called the "Perfect Cosmological Principle". An intense controversy developed between proponents of different theories of the universe, as discussed in (Kragh 1996).

[32] In his article, Kafka also mentioned having used a method programmed on a computer and announced that details would be provided in an internal report of the Institute for Astrophysics, in preparation (Quasars and Cosmology. Institutsbericht MPI-PAE/Astro 2/67). See also (Kafka 1968b). In 1968, a short paragraph entitled "Kosmologische Fragen und Quasars" was included in the Annual Report (Biermann and Lüst 1969, 87). In June, Biermann wrote to Kafka: "Thank you for the reprint of your article in Nature and the copy of your article "Quasars and Cosmology" which appeared as an institute report. Congratulations for the invitation to attend the summer school on astrophysics at Lincoln, Nebraska […] After you return, I would like you to tell me in more detail about the present position of the Burbidges on the question of the distance of the Quasars." Biermann to Kafka, June 6, 1967, AMPG, NLB, No. 18. The issue of possible cosmological interpretation of the red shift of quasars and the counting of quasars was also discussed in a draft of a letter from Biermann to W. Mattig, who had sent him his article on the subject (Biermann to W. Mattig, January 21, 1969, AMPG, NLB, No. 19).

the Institute for Astrophysics was involved in such debates.[33] He had been invited as one of the commentators at the end of the first day (dedicated to quasi-stellar radio sources) of the Third Texas Symposium held in Dallas in January 1967 and on that occasion had the opportunity to discuss the problem of the distribution of quasars in the Universe and observational cosmology with all the participants involved in this new field.[34]

## 4. Pulsars, black holes, and the possible evidence for the existence of gravitational waves

As we have seen, in bringing the suggestion that pulsars had to be identified super-dense stellar cores left behind after a supernova explosion, this breakthrough discovery provided the first definite proof of the existence of these highly compact stars—previously only theoretical entities—in which the central densities can be as high as $10^{18}$ kg/m$^3$, meaning that the effects of general relativity are strong.[35] This radically widened the perspective, firmly establishing the belief that strong gravitational fields may be of key importance for quasars, for violent events in the nuclei of galaxies, for supernova explosions and remnants, for the death by collapse of very massive stars and, in general, for the very compact astrophysical objects that were beginning to populate the universe of the 1960s. Toward the end of the decade, black holes, exotic objects having hitherto only a purely theoretical status, became serious—albeit much debated—astrophysical hypotheses. The discovery of pulsars did settle the existence of neutron stars as endpoints of stellar evolution of massive stars, and had the effect that "rather less was heard about the inherent absurdity of the more radical endstate, especially after Wheeler had dignified it with a name: 'black hole'" (Israel 2000). The longstanding commitment at the Max Planck Institute for Astrophysics to study the structure and evolution of stars, also performed with computer simulations, developed into research on very dense stars such as white dwarfs or neutron stars.[36]

---

[33] See for example (Doroshkevich et al. 1970; Longair 1971) for a discussion on such topics and related literature.

[34] Kafka to Biermann, February 27, 1967. AMPG, NLB, No. 18. In this long letter, Kafka gave a very detailed report on the conference, and Biermann answered that he himself would be at the University of Texas in a couple of weeks, hoping "to learn there something more about the present state of relativistic astrophysics and of the problems of the quasi-stellar radio sources." Biermann to Kafka, March 13, 1967, AMPG, NLB, No. 18.

[35] The large increase in the rate of publication of papers on the properties of neutron stars also included general relativistic aspects which were especially studied under the direction of John Wheeler at Princeton, who had also considered possible emission of gravitational radiation from spinning and vibrating neutron stars: "The radiations include neutrinos, X rays, long-wavelength electromagnetic waves, and gravitational waves" (Wheeler 1966, 393). For a review on neutron stars as of the end of the 1960s see (Cameron 1970). A discussion on the emission of gravitational radiation can be found on pp. 202-203.

During the 1960s, the epoch of X-ray astronomy was also beginning. According to Zeldovich's estimations, the shock wave originating when the gas surrounding a neutron star falls onto its surface should produce radiation primarily in the X-ray range. Moreover, plasma oscillations might arise in this zone (Zeldovich and Shakura 1969). The discovery of the first radio pulsars, which turned to be strongly magnetized neutron stars, had started a new bonanza of radio astronomy. In the same 1960s scenario, gamma- and X-ray astronomy activities were in progress at the Institute for Extraterrestrial Physics in the Munich suburb of Garching, while the Max Planck Institute for Astronomy in Heidelberg and the Max Planck Institute for Radio Astronomy in Bonn were both finally founded. New conditions for the interaction between nuclear physics, astrophysics, cosmology, optical and new astronomies were being created, widening the scope and context of what was being relabeled as the field of "cosmic physics". Scientists in the 1960s were beginning to look at the universe with the most diverse eyes, ranging from the large mirror of the Hale Telescope at Palomar Observatory to a tank filled with thousands of liters of dry-cleaning fluid buried deep underground capturing solar neutrinos, to arrays of detectors in the desert hunting for high-energy cosmic rays or, key catalyst of subsequent developments in this work, a swinging aluminum bar in Maryland, waiting for gravitational waves.

According to Einstein's theory of general relativity, accelerated masses produce gravitational waves, which propagate at the speed of light through the universe. The existence and physical properties of gravitational radiation became central to various research agendas as one of the important open questions addressed by the general relativity and gravitation community emerging from the mid-1950s onward, when "the availability of appropriate notions of what a gravitational wave is allowed physicists to put forward heuristic arguments for their existence and detectability" (Blum et al. 2018, 534).

---

[36] In July 1968, Kippenhahn wrote to Biermann, referring to white dwarfs, collapsing stars, binary systems, and mentioning the problem that for the study of such complex related phenomena one needed a more powerful computing machine and that they had further perfected their program on the evolution of stars, being at the forefront compared with other groups (Kippenhahn to Biermann, July 10, 1968, AMPG, NLB, No. 18). In 1969, Biermann himself lectured on neutron stars in the U.S. and was invited to talk on pulsars in more than one occasion, in particular in Italy, at the Scuola Normale Superiore in Pisa and at the Accademia dei Lincei in Rome (Luigi Radicati to Biermann, March 14, 1969, AMPG, NLB, No. 31). Kafka, too, was invited to lecture on the subject in the symposium "Pulsars, and High-Energy Activity in Supernova Remnants" held in Rome in which Biermann again gave a talk on related topics (Biermann and Lüst 1970). See also (Biermann 1969). In June, Wolfgang Kundt, one of Jordan's former students, like Ehlers and Schücking, wrote a letter to Biermann about Robert H. Dicke's gravity experiment on the oblateness of the sun, that is, its departure from a spherical mass distribution, which of course was highly interesting for Biermann (Wolfgang Kundt to L. Biermann, June 28, 1969, AMPG, NLB, No. 18).

In a summary of the Chapel Hill Conference on the Role of Gravitation in Physics, held in 1957, which was published in *Reviews of Modern Physics*, Peter Bergmann had expressed the following opinion: "[…] the most important nonquantum problem that has been discussed at this conference is the existence of gravitational waves." He added that their existence and properties "represent an issue of preeminent physical significance" (Bergmann 1957, 352–353). However, in the concluding summary published in the proceedings of the conference, Bergmann also remarked that the detection of gravitational waves would be an experiment "which is apparently not feasible, and is not going to be feasible for a long time." He further pointed out that there was no general agreement at the time about the existence of gravitational waves, a most important question still to be settled.[37]

One of the protagonists of this revival of interest was Felix Pirani, who studied with Alfred Schild and obtained his second PhD in physics at Cambridge University under Hermann Bondi.[38] In 1957, Pirani published what was to become an influential paper on gravitational radiation (Pirani 1957),[39] which, together with later contributions by Bondi, Ivor Robinson, and himself, as well as by Andrzej Trautman, overcame theoretical obstacles concerning the existence of gravitational waves and gave the "green light" to gravitational wave search (Hill and Nurowski 2017).[40]

---

[37] The proceedings of this conference, which was of great historical significance in the process of the renaissance of general relativity, have been republished as an open access volume (Rickles and DeWitt 2011).

[38] Felix Pirani: Interview by Dean Rickles, June 23, 2011. Transcript, AIP, https://www.aip.org/history-programs/niels-bohr-library/oral-histories/34463, accessed 30/7/2019. Bondi himself, who is known in particular for his work on cosmology, believed his most important scientific work was that on the theory of gravitation, specifically on gravitational waves (Hermann Bondi: Interview by David DeVorkin, March 20, 1978. Transcript, AIP, https://www.aip.org/history-programs/niels-bohr-library/oral-histories/4519, accessed 30/7/2019. He wrote a considerable number of papers on this subject in the period from 1957 to 1967, see in particular an article of 1957 in which he expressed the opinion that, contrarily to his own previous belief, "true gravitational waves do in fact exist. Moreover […] these waves carry energy." Bondi also devised a "primitive detector" for gravitational waves (Bondi 1957). A discussion of gravitational waves in the context of the renaissance of general relativity can be found in (Blum et al. 2016). A specific analysis is in the more recent article (Blum et al. 2018).

[39] Pirani's interest in the problem of gravitational radiation aroused during the Bern conference of 1955, marking the 50th anniversary of special relativity, also prompted Hermann Bondi to take up the problem. Pirani presented his new work on wave theory at the Chapel Hill Conference, when a lively discussion took place during the session on gravitational radiation (Kennefick et al. 1999, 215).

[40] For a longer, more detailed version, see (Hill and Nurowski 2016).

After the Chapel Hill conference, gravitational radiation became a key focus of theoretical studies in general relativity. In the meantime, Joseph Weber at the University of Maryland had chosen to spend his first sabbatical year (1955-1956) at the Institute for Advanced Study in Princeton, with John Wheeler and Robert Oppenheimer as his advisors (Trimble 2017, 265). He spent the second part of that academic year with Wheeler at the Lorentz Institute for Theoretical Physics in Leiden, which resulted in an article on gravitational waves that they co-authored, in which they addressed their reality and which was presented at Chapel Hill (Weber and Wheeler 1957). Encouraged by Wheeler himself, Weber accepted the challenge and pioneered the quest for the experimental detection of gravitational waves from astronomical sources.[41] In his first article presenting his views about the detection of gravitational waves, Weber thanked Pirani, Bergmann, and Wheeler for "helpful criticism" and acknowledged discussions with Robert H. Dicke (Weber 1960). Later, he spent the year 1962-1963 at the Institute for Advanced Study in Princeton and discussed the idea of a search for gravitational radiation with Freeman Dyson and Robert Oppenheimer. Both gave him strong encouragement (Weber 1980, 454).

Weber's experimental program was thus deeply embedded in the radical transformation characterizing the process of the renaissance of general relativity in the post-World War II period and in the related reorganization of knowledge. For several years, however, his experiments remained an isolated example. As underlined by Peter Saulson, "Weber's very concise discussion is remarkable for the prescience with which it foreshadowed not only his own work, but that of so many others. It also marks a watershed in the history of general relativity. In a single blow, Weber wrested consideration of gravitational waves from theorists concerned about issues such as exact solutions, and appropriated the subject instead for experimentalists trained in issues of radio engineering. The boldness and brilliance of this move are remarkable" (Saulson 1998).[42] However, Pirani himself—and certainly many others shared his opinion—was skeptical about the real possibility of detecting them: "The weakness of the gravitational interaction makes it exceedingly unlikely that gravitational radiation will ever be the subject of direct observation" (Pirani 1962, 199).

---

[41] As Wheeler himself recalled: "I gave such a feeling of reality to gravitational waves that Joe Weber has devoted himself since then to trying to detect gravitational waves." John Archibald Wheeler: Interview by Kenneth W. Ford, Session XI, March 4, 1994. Transcript, AIP, https://www.aip.org/history-programs/niels-bohr-library/oral-histories/5908-9, accessed 30/7/2019. Conversely, to have Weber as colleague during his Guggenheim fellowship first at Princeton and then at Leiden "was a real stimulus" to Wheeler. Session XII, March 28, 1994. Transcript, AIP, https://www.aip.org/history-programs/niels-bohr-library/oral-histories/5908-12, accessed 30/7/2019. A discussion on views about the existence of gravitational waves can be found in (Trimble 2017).

[42] See also (Levine 2004).

In his early papers, Weber speculated about the possibility of generating detectable gravitational waves in the laboratory but recognized that the chances of success were very low. As for the expected astrophysical sources, at the end of his 1960 paper, he had only briefly remarked that "The detectors which have been proposed are sufficiently good to search for interstellar gravitational radiation" (Weber 1960, 313). He later also mentioned as possible sources "events which might be associated with supernovae, neutron stars or closely spaced binaries"—again in the concluding lines (Weber 1963, 934).

Weber's work, in turn, inspired interest in such astrophysical objects as possible sources of gravitational waves. In his pioneering article, written in 1962, before the discovery of pulsars (and therefore of neutron stars), Freeman Dyson speculated that the usual formula giving the gravitational-wave energy flux from a binary star, leads, in the extreme relativistic case of a close binary collapsing system formed from a pair of neutron stars, to the prediction of a huge output of radiation. The powerful burst of gravitational waves—"the death cry of a binary neutron star"—should be detectable by Weber's existing equipment (Dyson 1963, 119).[43] This remark gave an extra stimulus to the pioneering experimental work of Weber, also prompting the physics and astrophysics communities to consider gravitational radiation—whose physical reality was becoming plausible—as a phenomenon of great potential importance in the physical world.[44]

The astrophysical scenario was thus coming up with very promising sources of gravitational waves. Back at the first Texas conference in 1963, when quasars had just been discovered, some theorists were suggesting that gravitational energy, released by a supermassive object, was responsible for the powerful radiation emitted (Rees 1998, 81). Similarly, gravitational waves had been proposed as a mode of energy loss by Hoyle and Fowler, and by Hoyle in a further publication (Fowler and Hoyle 1963; Hoyle 1963). In any case, as astrophysical

---

[43] The article was submitted as prize essay to the Gravity Foundation in April 1962.

[44] The renewal of interest in massive stars had been kindled by the already mentioned Fowler and Hoyle's suggestion that stars with mass of order of about $10^8$ solar masses might accumulate at the center of galaxies or in intergalactic space serving as the source of the "prodigious" energies involved in emission or storage in the radio galaxies and stars. After the discovery of quasars this led to the organization of the first Texas Conference and to discussions of the energy release in collapse of the core of a massive star and the related possibility of emission of gravitational radiation (Robinson et al. 1965).

theory—and computer simulations—began to reveal the characteristics of compact objects, such as neutron stars, attention was turned to the possibility of *detecting* the gravitational radiation emitted by gravitational collapse. In parallel with Freeman Dyson's speculations, the idea that close double stars with one white-dwarf component could radiate enough gravitational power to be astrophysically significant and even detectable—thus becoming "of interest as a test for the existence of gravitational waves"—was also suggested by others (Kraft et al. 1962, 314). In 1963, a mathematical expression for cosmic gravitational waves from *realistic sources* was given by Philip Peters and Jon Mathews, who worked out the gravity-wave emission from Newtonian binary star systems in bounded Keplerian motion (Peters and Mathews 1963). In 1964, when neutron stars were still a hypothesis, Hong-Yee Chiu, credited with coining the term *quasar* (Chiu 1964b), discussed a picture in which every supernova would "inevitably become a neutron star" ("the only way a neutron star may be formed") stressing that the rotational energy would be "dissipated, during the collapse phase, into gravitational waves." If perfected to their expectation, instruments designed by Weber and his associates "should be able to detect such waves many galaxies away." Chiu also mentioned the possibility that such neutron stars could be detected "by extraterrestrial x-ray telescopes" also adding that if detected, they would "pose interesting questions on our present theory of fundamental particles" (Chiu 1964a, 405). According to John Wheeler, too, the super-dense core remaining after a supernova explosion—together with neutrinos, X-rays, and electromagnetic radiation—might also emit gravitational waves (Wheeler 1966). Gravitational radiation was being considered during the 1960s also as a possible mechanism for both the dissipation and transfer of energy in the domain of relativistic astrophysics (Braginskii 1966).[45]

Spinning compact objects, too, were candidate sources of gravitational waves. In 1965, Chao-wen Chin, inspired by Chiu, discussed the gravitational radiation from a spinning body and used his calculation to estimate the energy-loss rate of a spinning collapsing neutron star (Chin 1965). In 1967, Franco Pacini pointed out that a spinning neutron star with a large magnetic field *would emit electromagnetic waves*, and might even be a source of gravitational waves (Pacini 1967), an hypothesis explicitly discussed by Wa-Yin Chau (Chau 1967). Such theoretical premises further clarify how the actual discovery of the pulsars, rapidly rotating neutron stars emitting a beam of electromagnetic radiation at very regular intervals announced in February 1968, was really instrumental in arousing considerable interest in the theory of very dense stars and gravitational collapse. As Pulsars were also quickly recognized as promising sources of detectable gravitational waves, Weber immediately estimated the expected fluxes of gravitational radiation from such objects and proposed a search on a specific band, suggesting that for this search he could modify his apparatus (Weber 1968a).[46] The possibility of the emission of gravitational radiation by such astrophysical sources attracted wider attention to Weber's ongoing efforts and, in 1968, he was asked to write a review article for *Physics Today*. The front cover of its April issue also featured a schematic representation of his experimental set-up (Weber 1968b).

---

[45] During the First Texas Conference Fowler proposed the energy transfer via gravitational radiation from the binary core of two collapsed stars as a mechanism for polar explosion of the envelope leading to quasars (Fowler 1964). His model was revisited in a more detailed analysis in (Cooperstock 1967).

Moreover, mid-June 1969 saw the publication of Weber's famous article claiming to have observed coincidences on gravitational radiation detectors based on resonating metal bars separated by a distance of about 1,000 km at Argonne National Laboratory near Chicago and at the University of Maryland: "There is good evidence that gravitational radiation has been discovered" (Weber 1969, 1324).[47] The announcement immediately spurred a wide interest: for example, the front cover of the issue of *Science News* dated 26 December 1970 was dedicated to "Black holes and gravity waves," also featuring the drawing of a vortex representing the black hole: "Because gravity-wave signals have actually been reported and because looking for them has become a good deal more popular than it used to be, theorists have been taking a close look at the kinds of objects proposed as possible sources to see which of them might actually be detectable." The article mentioned Ruffini and Wheeler and, of course, Weber, but it also reported William O. Hamilton's opinion about the efforts to detect gravitational waves at Louisiana State University, using cryogenic detectors to reduce the thermal noise and increase the sensitivity: "As long as supernovas were thought to be the only source of gravity waves whose signals were likely to be observable, it was not worth expending the money and the engineering effort to build something that might wait around 40 years before recording a signal. Now that Dr Weber's uncooled detectors are seeing gravity-wave events on the order of once a month, the need for the more sensitive detectors is evident" (Thomsen 1970, 481).

---

[46] Weber also suggested the Earth as a possible detector: because of its large mass, it has a very large cross section, which would make it absorb more gravitational waves. Freeman Dyson also studied the seismic response of the Earth to gravitational waves at pulsar frequencies (Dyson 1969). In the concluding lines, Dyson commented: "[…] we should remember the history of radio astronomy, which was greatly hampered in its early stages by theoretical estimates predicting that few detectable sources should exist. The predictions were wrong because the majority of sources were objects unknown to optical astronomers at that time. Whenever a new channel of observation of the Universe is opened, we should expect to see something unexpected. For this reason above all, the seismic detection of pulsars is not as hopeless an enterprise as the calculations here reported would make it appear." Scientists looking for gravitationally induced vibrations in Earth are mentioned in (Collins 2004, 116).

[47] Weber specified, "My definition of a coincidence is that the rectified outputs of two or more detectors cross a given threshold in the positive direction within a specified time interval. For the present experiments the time interval was 0.44 seconds. The magnitudes of the outputs at a coincident crossing enable computation of the probability that the coincidence was accidental. Observation of a number of coincidences with low probability of occurring statistically establishes, with good confidence, that the detectors are being excited by a common source. We may conclude that such coincidences are due to gravitational radiation if we are certain that other effects such as seismic and electromagnetic disturbances are not exciting the detectors" (Weber 1969, 1320). In this article, Weber acknowledged discussions, among the others, with C. W. Misner, H. S. Zapolsky, R. H. Dicke, F. J. Dyson, J. A. Wheeler, and P. G. Bergmann. See also Weber's preliminary description of the new series of experiments involving two detectors spaced about two kilometers apart, and the announcement that he had already observed a number of coincident events (Weber 1968c).

Soon, gravitational waves—as well as hard X-rays and gamma rays—would be envisaged by John Wheeler and Remo Ruffini as one of the most promising ways to detect black holes (Ruffini and Wheeler 1971).[48] In 1970 Franck Zerilli analyzed the problem of the pulse of gravitational radiation given off when a star falls into a "black hole" (Zerilli 1970), and Stephen Hawking's prescient article of 1971 even discussed gravitational radiation resulting from the collision of two black holes (Hawking 1971).

## 5. The impact of Weber's announcements at the Institute for Astrophysics

Joseph Weber's announcement caused a sensation in the physics community. The Max Planck Institute for Astrophysics quickly reacted to the new exciting perspective opened by his claims. His article was published in the June 16 issue of *Physical Review Letters* and by July there was a telephone conversation between Weber and Biermann, who was at the time in the U.S., where he was a regular visitor every year.[49] During the call, Biermann expressed the keen interest of his group in Weber's experiments, which he had most probably discussed with his collaborators immediately before leaving Munich.[50]

---

[48] Gravitational waveforms emitted by test bodies falling radially into a Schwarzschild black hole were given for the first time in (Davis et al. 1972).

[49] This is very clear from the bulk of Biermann's correspondence. In general, he traveled very often and thus, when he was far from Munich, he exchanged letters with his collaborators which are a precious source of information on the activity and movements of the group.

[50] Biermann to Weber, March 19, 1970, AMPG, NLB, No. 48. A telephone call was made between Aspen, Colorado, where Weber often spent time (as acknowledged in his articles) and Boulder, where Biermann had

Shortly afterwards, Peter Kafka reacted to Weber's article with a detailed analysis of the possible sources of gravitational waves, also discussing related difficulties of interpretation of his data, concluding that "All possibilities to explain the large number of events observed by Weber seem rather unlikely and demand more or less 'accidental' sources" (Kafka 1969, p. 138; see also: Kafka 1970a, 1970b). Signals of the magnitude and rate observed by Weber were not easily explained on the basis of known astrophysical phenomena. If they originated from a source near the center of our Galaxy, as he suggested, it appeared rather hard to reconcile the energy fluxes implied with other estimates of rate of energy loss by the Galaxy, as discussed for example by Dennis Sciama (Sciama 1969) or by Kafka himself in the essay "Are Weber's Pulses Illegal?", submitted to the Gravity Research Foundation competition.[51]

The characteristics of Weber's gravitational wave antennae were immediately studied by Hermann Ulrich Schmidt, while Kafka explored in detail the possible consequences of the gravitational waves "supposedly discovered by Weber."[52] Gerhard Börner, who had been Heisenberg's and Hans P. Dürr's PhD student at the Ludwig Maximilian University of Munich, with a dissertation on quantum field theory in cosmology (*Feldtheorien im de Sitter-Raum unter besonderer Berücksichtigung der nichtlinearen Spinortheorie*),[53] was now working at the Institute for Astrophysics on relativistic cosmology and models of neutron stars.[54] Börner also collaborated on theoretical models for super-dense matter with Hans A. Bethe and Katsuhiko

---

spent the months of July and August in 1969 giving lectures (Biermann and Lüst 1970, 79).

[51] Kafka's essay was awarded the second prize for the year 1972 of the annual award offered by the Gravity Research Foundation (Anonymous 1972).

[52] These research activities, together with Biermann's studies on some characteristics of the density of pulsars were announced in the new section of the Annual Report for 1969 entitled "Relativistische Astrophysik, Quasare und Pulsare" (Biermann and Lüst 1970, 86–87).

[53] See related publications (Börner and Dürr 1969, 1970; Börner 1970) and Heisenberg's parallel interests during the 1960s in cosmological problems, unified theory of elementary particles and non-linear spinor theory (AMPG, III. Abt., Rep. 93, No. 913, 950, 951, 953, 954, 961, 965, 966, 982, 999). Heisenberg's non-linear spinor theory is a main focus of a detailed historical account of his quest towards a theory of everything during the 1950s (Blum 2019).

[54] See for example (Börner 1973a, 1973b). At the end of July 1969, Börner made a long report to Biermann about the Enrico Fermi summer school in Varenna dedicated to general relativity and cosmology, directed by Rainer K. Sachs, a former student of Peter Bergmann's at Syracuse, to which Weber, too, participated. Börner mentions discussions there about Weber's claims that the gravitational waves he was detecting apparently originated from the center of our galaxy (G. Börner to L. Biermann, July 30, 1969, AMPG, NLB, No. 18). On August 6, Biermann confirmed that he had got a message from Weber indicating that the gravitational radiation measured using his instruments was in fact not isotropic. Biermann also asked Börner whether he had taken a look at "less high densities" and mentioned that he was handling the problem of pulsars in connection with an invitation he had received at a symposium on the subject organized by the Accademia dei Lincei in Rome.

Sato, who had been guests in Munich (Biermann and Lüst 1971, 91; Bethe et al. 1970; Börner and Sato 1971) and later spent a long period of time in Kyoto, at the Research Institute for Fundamental Physics directed by Hideki Yukawa, as well as in the U.S. With Börner, cosmology later became one of the fields included in the research agenda of the Max Planck Institute for Astrophysics.[55]

By early summer of 1969, both Biermann and Heisenberg were working towards intensifying research on gravitation theory and relativistic astrophysics.[56] In this regard, they shared the common aim to invite the relativist Jürgen Ehlers to spend a long period of time at their Max Planck Institute. Ehlers, who had a professorship at the University of Texas, Austin, was now holding visiting professorships in Germany (Allen et al. 2008).[57] It became Biermann and Heisenberg's ambition to have him back in Germany. In spring 1969, Biermann had met Ehlers in Göttingen proposing him to spend a few weeks in Munich in October, at the Institute for Physics and Astrophysics, during which time Ehlers might discuss gravitation theory and relativistic astrophysics topics at different occasions at the institute.[58] At the time, Ehlers, had

---

[55] A specific section dedicated to cosmology appeared in annual reports from 1986 onward (Hillebrandt and Schmidt 1987, 205).

[56] See minutes related to the meeting of 9 June 1969 of the search commission for Heisenberg's succession in AMPG, II. Abt., Rep. 62, No. 437, Fol. 273.

[57] Jordan had even favored Ehlers as his own successor in Hamburg (see related correspondence between Jordan and Heisenberg during winter 1967-1968 in Heisenberg's papers, AMPG, III. Abt. Rep. 93, No. 1745). In spring 1969, Ehlers had emerged as a possible candidate to Heisenberg's succession, when a decision had not yet been taken by the search committee whether a theoretical or an experimental physicist should lead the Institute for Physics after Heisenberg's retirement (see documents in AMPG, II. Abt., Rep. 62, No. 437). During discussions about the possibility of appointing a theoretician, in particular an expert in general relativity, it was also mentioned that in Germany Einstein's theory had "somewhat receded into the background" at universities, something that Jordan had pointed out in several occasions. In May 1969, Gentner and Jordan had an exchange of correspondence on this question, and Ehlers' name was definitely the most favoured according to the opinion of several relativists (Gentner to Jordan, 13 May 1969 and Jordan to Gentner 19 May 1969, AMPG, II. Abt., Rep. 62, No. 437, Fol. 48-59).

[58] Agreements about Ehlers' stay in Munich were made in June of that year. Following a meeting in Göttingen a short time previously, Biermann proposed him a sojourn of a few weeks in October (L. Biermann to J. Ehlers, June 16, 1969, AMPG, NLB, No. 18). This letter is followed by a draft of a letter not sent, probably prepared before their meeting in Göttingen (see handwritten note mentioning Göttingen) that is interesting because of its more detailed content, which illustrates the kind of research questions they would have liked to address at the institute [for Physics and Astrophysics] and the related idea of having Ehlers for a long period of time: "In connection with the most recent observations in the field of relativistic astrophysics, we have asked ourselves the question of how to make the theory of gravitation and the related questions of cosmology—also in the context of the more recent observations of 3° K radiation—and the pulsars, the subject of works here at the Institute. The Institute for Astrophysics already has an old tradition in some areas of mathematical physics and applied mathematics, and the Institute for Physics has long been interested in the relationships between quantum field theory and gravitation. In this context, the question has arisen as to whether we could hire you for at least one or two years for our institute [translation by the authors]." The letter concluded with several hypotheses on different options for the

just published a wide overview on the state of cosmology in relationship with the impact of the recent discoveries of quasars, pulsars, the cosmic microwave background, which were pushing general relativity to the forefront, together with systematic experimental efforts to understand its observable predictions (Ehlers 1969).[59] As Jordan himself had stressed in a letter to Gentner, general relativity in the Federal Republic of Germany was not sufficiently appreciated "by the physical colleagues" who seemed to have the idea "that the theoretic work in this direction still has a similar character today, as at the time of the great wave of speculation inspired by Weyl's time, which were increasingly lost in the areas of the empirically untenable." He further added that, "In reality, the *present* theoretic treatment of this area is very much geared towards serving the astrophysical applications (e.g. gravitational collapse) and the novel tests made possible by satellite technology and radar technology. Today, several working groups are busy putting the most modern technological and experimental possibilities into the service of new empirical examinations of the theory." As a last remark, Jordan mentioned that "in Canada alone more than 25 theoreticians belong to the active researchers in the field of Einstein's gravitational theory" and concluded that he was convinced that it was "urgent to end the self-exclusion of the Federal Republic of Germany from this so current modern field of research." Heisenberg, and even Born, according to Jordan, fully shared his worries and the need for an urgent solution to such situation.[60]

In early October of that year, Ehlers was in Munich (Biermann and Lüst 1970, 79). The recent discovery of pulsars, as fast-rotating super-dense neutron stars, had given rise to a series of new perspectives which were of the utmost interest for exploring connections between general relativity and astrophysics, a subject Biermann had the opportunity to discuss with Ehlers.

---

two scientists to see each other in Germany or in the U.S., where Biermann was going for 2 months starting from July 11.

[59] See also his contribution on gravitational waves in the lectures at the summer school on general relativity organized by the well-known Italian relativist Carlo Cattaneo (Ehlers 2011). Ehlers had also recently edited a series of three volumes containing the proceedings of a summer seminar held at Cornell University in 1965 on problems in relativity and astrophysics, which included lessons on theoretical developments in general relativity, experimental tests of general relativity, stellar structure and gravitational collapse, gravitational radiation, observational cosmology, and cosmic rays (Ehlers 1967).

[60] Jordan to Gentner, 19 May 1969, AMPG, II. Abt., Rep. 62, No. 437, Fol. 48.

From then on, things moved quickly. Biermann proposed that Ehlers should move to the Max Planck Institute for Astrophysics[61] and, at the end of October, Heisenberg and Biermann sent a joint letter to Adolf Butenandt, then President of the Max Planck Society, in which they emphasized how during the last year general relativity and gravitation question had become relevant both at the Institute for Astrophysics and for Physics, especially in relationship with gravitational waves and neutron stars. For this reason, the Munich institutes would strongly benefit from the presence of a renowned relativist like Jürgen Ehlers.[62] In documents related to Ehlers call to Munich it is clearly stated how *both* Heisenberg's and Biermann's scientific interests would benefit from having Ehlers at the institute. A main aim was also to build a bridge between unified field theory and gravitation theory in connection with new related interests in astrophysics and the idea of creating a group working on gravitational wave experiments. This would thus also create a deeper relationship between theory and experiment.[63] A first step in this direction would be to call Jürgen Ehlers and open new perspectives at the institute in interdisciplinary studies encompassing astrophysics and theoretical work on the unified field theory. On 7 November 1969, a commission to appoint Ehlers as a scientific member of the Institute for Astrophysics was formed.[64] On the following 3 March 1970, the Senate confirmed the appointment, remarking that Ehlers' visit to Munich had shown that his presence would be of the greatest importance both for the Institute for Physics (Hans P. Dürr's theoretical group) and Biermann's Institute for Astrophysics, as well as for the Institute for Extraterrestrial Physics. Together with the Max Planck Institute for Plasma Physics led by Schlüter, all these had been born from Heisenberg's Institute for Physics established in Göttingen after the war, as a continuation of the Kaiser Wilhelm Institute for Physics founded in 1917 in Berlin.[65]

---

[61] Minutes of the 15th meeting of the board of trustees (Kuratorium) of the Max Planck Institute for Physics and Astrophysics, 17.03.1970, AMPG, II. Abt., Rep. 66, No. 3069.

[62] L. Biermann and W. Heisenberg to Adolf Butenandt, October 31, 1969, AMPG, III. Abt., Rep. 93, No. 1667.

[63] Minutes of the 15th meeting of the board of trustees (Kuratorium) of the Max Planck Institute for Physics and Astrophysics, 17.03.1970, AMPG, II. Abt., Rep. 66, No. 3069.

[64] The same committee was also involved in Heisenberg's succession. Such appointment proceedings are described in reports of the committee to the chemical-physical-technological section of the scientific council of the Max Planck Society which are part of Rep. 62. From now on these are shortened to "CPTS minutes" + dates/numbers. (CPTS meeting minutes of 07.11.1969, AMPG, II. Abt., Rep. 62, No. 1757).

[65] A department for Astrophysics led by Biermann existed since 1948 within the Max Planck Institute for Physics, becoming a sub-institute in 1958. The Institute for Extraterrestrial Physics led by Reimar Lüst was founded in 1963 as a new sub-institute, while the Institute for Plasma physics was founded in 1960 as an autonomous entity, from

Connections between gravitation theory and topics such as gravitational waves, neutron stars, quasi-stellar systems, and also quantum field theory and future attempts to detect gravitational waves had come into focus as key research questions. Ehlers' arrival would not change the overall organization: research on elementary particle physics would continue as in the past, but general relativity and gravitation would create a new link with astrophysics. With the incursion into gravitational wave experiments (see below), the Institute for Astrophysics would also move into experimental astrophysics based on a strong theoretical standpoint, a process that was characteristic of the Munich family of institutes.

Ehlers became a scientific member of the Institute for Astrophysics as of 1 June 1971.[66] At that time, general relativity was becoming a major branch of physics, also boosted by the newly established fields of relativistic astrophysics and observational cosmology (Blum et al. 2018).

## 6. Heinz Billing's resonant bar experiments and the promise of gravitational-wave astronomy in Munich

In the meantime, in late November 1970, the possibility of starting a gravitational wave experiment was being seriously considered by Biermann's group.[67] The section headed "Relativistische Astrophysik, Quasare und Pulsare" of the 1970 Annual Report was now clearly showing the establishment of a new research line in which, for the first time, the "preparation of a gravitational-wave experiment" is mentioned, together with studies on matter at supra-nuclear density and neutron star models.[68] The extreme physics of these stars could allow

---

which the Institute of Quantum Optics—having a relevant role in this story—was born in 1979. As we will see in the following pages, the family will further enlarge with the foundation of two new Max Planck Institutes related to general relativity and gravitational-wave research.

[66] On February 9, 1971, during the meeting of the CPT Section of the Scientific Council, it was communicated that Ehlers had accepted, and that he would take up his position on June 1, 1971 (AMPG, II. Abt., Rep. 62, No. 1761).

[67] See Biermann, to Börner, November 26, 1970, AMPG, NLB, No. 20. On November 26, 1970, Hermann Ulrich Schmidt, who was spending some time at the National Solar Observatory at Sacramento Peak in New Mexico, was writing to Biermann about discussion he was having with Ehlers, Weber and Remo Ruffini about beginning a gravitational wave experiment in Munich. See also answer from Biermann on December 8 (Schmidt to Biermann, November 26, 1970, and Biermann to Schmidt, December 8, 1970, AMPG, NLB, No. 21).

[68] The group included Kafka, Friedrich Meyer, and Gerhard Börner (Biermann and Lüst 1971, 91). See also works published during the initial phase, as (Kafka and Wills 1972). Kafka stressed that the most powerful emission would come from "the collapse of a rotating star towards a black hole" or the "fusion of two black holes in a dense cluster" (Kafka 1970b, 436).

astronomers to probe physics in very strong gravitational fields—and general relativity would be the tool to understand the structure of such highly energetic, compact astrophysical objects.

In parallel with intense theoretical work on general relativity and relativistic astrophysics, plans for the gravitational waves experimental activity at the Max Planck Institute for Astrophysics continued, immediately involving on the experimental side Heinz Billing, who was leading the computing group since the early 1950s, now successfully returned to physics.[69] Wheels were put in motion and work began in earnest in 1971, when the gravitational wave experiment had its own specific section in the Annual Report.[70] The aim was "to confirm or disprove the existence of gravitational pulses suggested by Weber as an explanation of his results" (Billing et al. 1975, 111). With the arrival of Ehlers in June 1971, the new Department for Gravitation Theory and Relativistic Astrophysics was established.[71]

As the Munich setup was planned to be as close as possible to Weber's experiment, in January-February, both Billing and his new assistant, Walter Winkler, visited Weber at the University of Maryland in order to become familiar with his antenna and obtain all the information that would be useful for their future work.

---

[69] As recalled by Billing: "In 1970, I was surprisingly visited in my study by two Biermann's senior staff members, H.U. Schmidt and Friedrich Meyer, and asked if I would be willing to repeat Weber's gravitational wave experiment […] This task appealed to me immensely. A whole new field of research, working at the limit of the measurable and finally again practicing real experimental physics! But there was one difficulty. The only one who was free to do a new job in my department was me. All my employees were deep into important ongoing projects [our translation]." And so, having at disposal a couple of free positions, Billing hired a new assistant, Walter Winkler, through an announcement in *Die Zeit*, a most renowned weekly newspaper in Germany (Billing 1994, 156-161). Winkler began to work on January 1, 1971, and they soon went to visit Weber in the U.S. for a whole month.

[70] They specified that the decision to repeat Weber's gravitational-wave experiment had been taken both because of its great astrophysical significance and the still pending difficulties in evaluating Weber's findings. The prerequisites for this were particularly favorable at the institute, as the necessary engineering and electronic experiences were available at the Numerical Calculators Division, while the local astrophysicists would be able to handle the theory and the statistical problems, and the addition of Ehlers would guarantee the close connection with the general theory of relativity. It was further emphasized how Weber's detector could be improved (Biermann and Lüst 1972, 326).

[71] Ehlers carried on his fundamental work on the exact solutions of Einstein's theory, Börner continued his theoretical work on pulsars and neutron stars, also as supernova remnants. Kafka and Meyer analyzed Weber's evaluation methods in view of their own planned experiment. Martin Walker studied problems connected to the theory of black holes. Activities included the organization of a 2-day workshop on pulsars at the institute and participation in conferences related to relativistic astrophysics and general relativity and gravitation (Biermann and Lüst 1972). During the course of 1971, new members had joined Billing in the gravitational wave project: W. Winkler (from January 1), and John M. Stewart and Martin Walker (both from October 1). John Stewart was a former student of Dennis Sciama and George Ellis in Cambridge at the Department of Applied Mathematics and Theoretical Physics and at the Centre for Theoretical Cosmology. http://www.ctc.cam.ac.uk/news/161121_newsitem.php, accessed 1/9/2019. Both Stewart and Walker were meant to work as assistants to Ehlers (see related correspondence between Ehlers and Biermann between late 1970 and spring 1971, AMPG, NLB, No. 20).

But for a coincidence experiment they needed a second antenna, far from Munich. They were lucky, because, independently from them, a German colleague, the electronics engineer Karl Maischberger, and the physicist Donato Bramanti had also begun to work on a Weber-type gravitational wave antenna at the European Space Research Institute (ESRIN) in Frascati, near Rome, with which the institute had already interacted in the past years (Bramanti and Maischberger 1972).[72] A Conference on Cosmic Plasma Physics was organized by ESRIN in September 1971. Biermann participated in this conference and it was certainly an opportunity for him to become acquainted with local plans on the gravitational wave experiment and to establish a collaboration.[73] In the early 1970s, a second experimental activity for gravitational wave detection had also begun in Italy, at the University of Rome, led by Edoardo Amaldi and Guido Pizzella, which evolved along a different research line.[74]

The Munich resonant bar—a long aluminum cylinder reproducing Weber's setup that should ring at a certain frequency in response to a gravitational wave—began operating as of October 1972.[75] The aim was to test whether the pulses of gravitational radiation reported by Weber were detectable in coincidence between Munich and Frascati.[76] The first negative results, in conflict with Weber's, were presented in June 1973, in Paris, at the International Colloquium on Gravitational Waves and Radiation (Kafka 1974a).[77]

---

[72] While intending to be as close as possible to the original experiment, they still made several improvements, which made their detector—together with the similar one built in Frascati—"the most sensitive room-temperature bar experiment at that time" (Winkler 2018, 15).

[73] For several years, Biermann had been in contact with the astrophysicists Livio Gratton and Franco Pacini, working at the Laboratory for Space Physics in Frascati. In December 1970, Biermann wrote to Pacini: "I read your paper on 'Neutron stars, Pulsar Radiation and Supernova Remnants' which you had sent me earlier, with great interest." Biermann to F. Pacini, December 10, 1970, AMPG, NLB, No. 30. In February of the following year, Biermann mentioned his participation in the Conference on Cosmic Plasma Physics. Biermann to F. Pacini, February 3, 1971, AMPG, NLB, No. 30. The same folder also contains correspondence with Livio Gratton, who had initiated research on relativistic astrophysics at Sapienza University in Rome and was called from abroad by Edoardo Amaldi to the chair of astrophysics (Bonolis et al. 2017).

[74] The relationship between Biermann and Amaldi dated back to the end of the 1950s, when Enrico Persico and Amaldi were setting up at the Physics Department of Sapienza University in Rome the Laboratorio Gas Ionizzati, where research on plasma and thermonuclear fusion was performed (see correspondence in Edoardo Amaldi Archives, Sapienza University of Rome, Enrico Persico papers, Box 16). See, also, for example, Edoardo Amaldi's letter to L. Biermann, June 17, 1959, thanking him for lessons held in Rome on plasma and cosmic rays. Biermann had offered that Rudolf Kippenhahn might stay in Rome for a few months, but Amaldi wanted someone who could remain for at least 2 years to give a strong support to initial activities on plasma research (AMPG, NLB, No. 28) (Bonolis 2012). On the other hand, Amaldi's interest towards general relativity and gravitational waves had developed since the end of the 1950s, leading him to encourage his post-doc student Remo Ruffini to visit Pascual Jordan in Hamburg and later supporting his application for an European Space Research Organization (ESRO) 2-year fellowship to be spent in the U.S., at Princeton with John Wheeler and at Maryland University with Weber. In July 1967, in a letter written to Reimar Lüst (who would soon become Vice President of ESRO) Amaldi mentioned his intention of setting up an experimental group working in the field of gravitational waves "at the return of Ruffini" (E. Amaldi to R. Lüst, July 7, 1967, Edoardo Amaldi Archives, Box 375, Folder 3). On the Roman activities see (Bonolis and La Rana 2017; La Rana and Milano 2017; Pizzella 2008). A more complete historical overview emphasizing experiments made by the Rome group can be found in (Pizzella 2016).

As a sign of the growing importance of its status, the Munich-Frascati gravitational wave experiment became the first research activity presented in the Annual Report 1973. No signal had been detected, but investigations were underway that aimed at understanding whether improvement of antennae could further extend the sensitivity.[78] In 1974, the experiment was still the first item presented in the Annual Report: it was considered to have reached the highest performance possible for that type of antennae, with the highest sensitivity among the coincidence detectors used (Pinkau 1975, 103).

Triggered by Weber's announcement, other groups had also started experiments to analyze and test Weber's results: in the United Kingdom,[79] in the United States at IBM Research Center in Yorktown Heights (New York), and Bell Laboratories in Holmdel (New Jersey), (Levine and Garwin 1973; Tyson 1973; Douglass et al. 1975), in Japan (Hirakawa and Narihara 1975), and in the Soviet Union, where discussions on gravitational wave detection began already around 1960. One of the main promoters of gravitation research in USSR was Dmitri Ivanenko (University of Moscow), who had just edited the volume *The Newest Problems of Gravitation*, where the Russian translation of papers on the problem of gravitational waves, notably by Bondi and Weber, were included.[80] Ivanenko, was a member of the International Committee on General Relativity and Gravitation established after the Royaumont conference of 1959 near Paris, which brought together scientists working on both sides of the Iron Curtain with the task of coordinating research activities and organization of international conferences (Lalli 2017, 51–53). At that time, Zeldovich was shifting his research interests from nuclear and particle physics to astrophysics, general relativity, cosmology, and the new astronomies. He was one of the first to think in terms of the Big Bang Universe as a natural laboratory for particle physics and to call the universe "a giant accelerator." For the last 25 years of his life, astrophysics and cosmology had a central place in his thinking, as well as in the thinking of his collaborators (Sakharov 1988). Zeldovich immediately recognized the importance of Weber's gravitational-wave experiments and their possible role as probes to explore fundamental physics and cosmology and as a tool in astronomy and astrophysics to study compact objects, and organized a meeting of his theoretical group with Vladimir B. Braginsky's experimental group, at Moscow State University.[81] Braginsky assembled a scientific group and was the first, after Weber, to build a resonant bar—of the same size as those of Weber—and while active theoretical studies continued, experimental efforts were performed from the late 1960s, repeating the search for coincident signals on separated Weber-type antennae (Braginskii et al. 1969). Results obtained were negative (Braginskii et al. 1972).[82]

---

[75] Both Munich and Frascati built detectors as close to Weber's as possible, including a close match with his resonant frequency of 1660 Hz (Bramanti et al. 1973).

[76] As ESRIN was going to be closed, the Frascati experiment would become part of the Munich program, but run in Frascati (Pinkau 1973, 101). See also (Kafka and Meyer 1972).

[77] At the same conference, Silvano Bonazzola and his group presented the results of the Weber-type detector built at the Meudon observatory, mentioning double and triple coincidence research with the Munich and Frascati detectors (Bonazzola et al. 1974). Kafka gave a talk with the same title at the Symposium "Gravitational Radiation and Gravitational Collapse" of the International Astronomical Union held in Warsaw in early September 1973 (Kafka 1974b). See also discussions during the Sixth Texas Symposium of 1972 (Weber et al. 1973).

[78] In 1973, the group included Billing, Kafka, Meyer, Lise Schnupp, and Winkler. Maischberger, too, was a member of the scientific staff at the institute (Pinkau 1974).

[79] The group in Glasgow was then led by Ron W. P. Drever, who later became a member of the team initially running the LIGO project. Interestingly, Drever's interests in those days went from the search for gamma rays from pulsars to radio signals associated with gravitational waves, searching specifically for pulses from the galactic center which might be correlated with the events reported by Weber (Charman et al. 1970, 1971). The Glasgow group (J. Hough, J. R. Pugh, R. Bland and R. W. P. Drever) used a slightly different type of detector and instead of a single bar, they had a system made by two separate aluminum bars with piezoelectric transducers cemented between them to monitor changes in their separation. This arrangement would give higher coupling between the mechanical and electrical systems, and so a much larger fraction of mechanical energy would be communicated to the transducers, obtaining a larger electrical output than in a Weber detector. For their observations two of these detectors were set up 50 m apart and results from data recorded between fall 1972 and April 1973 were submitted in early September: "In this time," they claimed, "we have observed one distinctive signal which fulfils the requirements expected of an event due to a short pulse of gravitational radiation". But they finally concluded that it was unlikely that the signals reported by Weber in 1970 were due to pulses of gravitational radiation of duration less than a few milliseconds, even if, based on their observations, they did not exclude "the possibility that Weber may have detected bursts of gravitational radiation of much longer duration" (Drever et al. 1973).

In December 1974, also the Glasgow group reported a negative result providing evidence "against the hypothesis that the signals reported by Weber are caused by a large flux of very small pulses" (Hough et al. 1975, 501).

---

[80] For an overview of experimental research on the detection of extraterrestrial gravitational radiation performed in the Soviet Union since the late 1960s, see (Rudenko 2017).

[81] Since the end of the 1950s, Braginsky had discussed the possibility of measuring the speed of propagation of gravitational waves (Braginskii et al. 1960), and in his review of 1966 (the Russian version was published in 1965), he also presented a list of the most noteworthy potential sources to be detected (Braginskii 1966).

[82] The authors reported that no statistically significant excess coincidences had been found in their experiment. See also (Braginskii et al. 1974).

Other searches were carried out by various groups, all with predominantly negative results. The Munich-Frascati experiment reported results of the first 150 days of coincident data 1975 (Billing et al. 1975)[83] and, in March 1976, after 580 days of total useful observation time, the detectors were dismantled and the experiment stopped, with the conclusion that "The most interesting aim of gravitational wave astronomy will be the observation of stellar collapse" and the hope that "the many new antennae, being developed now and during the next decade, will be able to detect a few gravitational collapse events per year, and thus provide most valuable information on extreme states of matter and final stages of stellar evolution which will otherwise remain hidden."[84] Billing himself thought that the negative outcome was "not tragic," and that they had built what for some time had been the most sensitive antennae worldwide. As Kippenhahn commented: "Billing and his group are those who up to now have been the best in *not* finding gravitational waves" (Billing 1994, 163).[85]

In the early 1970s, the Max Planck Institute for Astrophysics considerably expanded its research activities and the growing number of national and international visitors corresponded to a similar flux of internal members visiting scientific centers abroad or invited to give talks, as evident from the annual reports at the time.[86] This went in parallel with the explosive developments of astrophysics and cosmology, strongly supported by the rapidly evolving field of the new astronomies whose birth had been fueled by the advent of the space age. These new technological windows also promised to allow studies on astrophysical processes that only seemed possible within the framework of general relativity. For instance, black holes and the search for their observational evidence, theories of quasars, neutron stars, compact X-ray sources, the physics of high density and nuclear matter, and the distribution of quasars in the universe were becoming popular subjects addressed at conferences. In 1973, the 16th Solvay

---

[83] The Frascati-Munich group claimed to have "set the lowest limits so far obtained for the rate of incoming short gravitational pulses stronger than a few times $10^5$ erg/cm$^2$ Hz at frequencies around 1660 Hz." The same frequency band had been used by Weber. See (Billing and Winkler 1976, 665).

[84] It seemed "appropriate" to publish the final negative results because—they claimed—"our experiment was as similar to Weber's as possible, whereas all other coincidence experiments deviated in one way or the other [...] Moreover, we think we have set the lowest limits obtained by Weber-type experiments over a reasonably long period of observation." (Kafka and Schnupp 1978, 97).

[85] For a discussion on the negative results from the different groups and the response to Weber's claims, see (Saulson 1998).

[86] The guests visiting each year the Institute for Physics and Astrophysics, also including the Institute for Extraterrestrial physics, grew from around 30 at the beginning of the 1960s up to around 100 in the early 1970s.

Conference in Physics, entitled "Astrophysics and Gravitation," was promoted by Edoardo Amaldi (then President of the Scientific Committee), strongly supported by Biermann and Heisenberg.[87] At that time, when gamma and X-ray astronomy were already becoming a key tool for understanding the high-energy universe, the quest for the detection of gravitational waves using the new technique of laser interferometry was still in its infancy (see below), but under the term of "Gravoastronomie", the field was already contributing to set the stage for the emergence of multi-wave multi-messenger expansion within a cluster of institutes of the Max Planck Society (Kafka and Meyer 1972).[88]

"The future of gravitational-wave astronomy looks bright whether or not Weber is actually detecting gravitational radiation," remarked Press and Thorne as early as 1972. Such hopes were embedded in the wider awareness that the windows of observational astronomy had become broader, now including "along with photons from many decades of the electromagnetic spectrum, extraterrestrial 'artifacts' of other sorts: cosmic rays, meteorites, particles from the solar wind, samples of the lunar surface, and neutrinos" (Press and Thorne 1972, 335). After a few years, despite the failure of all experiments searching for gravitational waves, and despite the "apparent disagreement between the results of Weber's experiments and those of other workers" leading to a great amount of controversy, the status of the field appeared so well established that Ronald Drever could point out that "the consensus view that Weber's results are not due to gravitational radiation seems to me so likely to be correct that it is more profitable to concentrate now on development of detectors of very much greater sensitivity" (Drever 1977, 16).

By the completion of the "first-generation" detectors around 1975, the design and early development work for a "second generation" was already underway.

---

[87] See W. Heisenberg to E. Amaldi, April 16, 1973 and other correspondence related to the 1973 Solvay conference (AMPG, NLB, No. 30).

[88] Significant levels of gravity waves are produced by bulk motion of huge amounts of mass and transmitted almost undisturbed through all forms and amounts of intervening matter. Information carried by gravitational waves is thus complementary to the information carried by electromagnetic radiation.

## 7. The transition from resonant bars to laser interferometry: An 'original sin'

The first wave of experiments—as well as the discovery of pulsars and the longstanding aim to detect gravitational radiation pulses produced in catastrophic collapse of stars resulting in supernovas or black holes—had prompted other researchers to propose alternative detectors designed to search for such short pulses, claiming higher sensitivities than those of Weber's original experiments, but within about an order of magnitude of them. A first obvious approach involved using larger bars of aluminum—or new types of material—and cooling them down to very low temperature (2 K or less, near the absolute zero) to reduce thermal noise, measuring their oscillations by totally new types of mechanical/electrical transducers. Developments in this direction had been proceeding for several years at Stanford University, at Louisiana State University and at Sapienza University in Rome.[89] Another very challenging proposal from a technical point of view came from Braginsky's group at Moscow University: instead of bars they were experimenting the possibility of building relatively small gravitational wave detectors using single sapphire crystals weighing only a few kilograms, which should be very efficient in discriminating between thermal noise and gravitational wave pulses.[90]

While these groups were concentrating on the problem of reduction of the background effects, an alternative approach to improving sensitivity could be obtained increasing the displacement caused by the wave. As the variation of the distance between the test masses induced by the passage of a gravitational wave is proportional to the distance between the masses, one must increase the separation between the test masses as much as possible. It is this extremely small change in separation which has to be experimentally detected against a background of perturbing influences such as thermal and seismic vibrations.[91] The basic idea behind this new approach was to continually compare the lengths of the arms of optical interferometers by bouncing laser beams between pairs of mirrors at the ends of each arm, and then making the two beams converge on a point and overlap. In the absence of gravitational waves, the beams' electromagnetic oscillations cancel out. If there is a space-time disturbance caused by gravitational waves, the arms change length, and the laser beams no longer cancel each other out: light is detected.[92] Any relative distance changes in two optical paths at right angles to one another would detect displacements due to a gravitational wave propagating in a direction normal to the plane of the system. The advantage of this method is that the mirrors, acting as test masses, can be placed kilometers apart, so that a gravitational wave induces larger relative motions. The changes in optical path might be further increased by reflecting each beam back and forward many times between each pair of masses to enhance displacement sensitivity. But a most important feature of interferometer antennae—which are potentially more sensitive than resonant-bar detectors—is that they are *inherently broadband*, being also sensitive over a much wider range of frequencies than had been practicable with bar detectors, and can detect and measure the wave forms of all classes of sources. However, laser systems of course had also the disadvantage of being technologically more complex and, in particular, *more expensive* than bars.

---

[89] Since early 1971, a long-term project for a second-generation detector of this type, a cryogenic resonant bar detector, had been developed in Rome, at the Sapienza University, by Edoardo Amaldi and Guido Pizzella (Pizzella 2016); La Rana and Milano 2017; Bassan and La Rana 2017). William Fairbanks and William Hamilton at Stanford and Louisiana State University, respectively, and David Blair's team at the University of Western Australia in Perth were also developing cryogenic resonant bars cooled to liquid-helium temperatures, that is, about 270 degrees below zero on the Celsius scale. In 1975, studies on an interferometric detector were also initiated by Amaldi and Pizzella's group in two dissertations by Massimo Bassan and Livio Narici, a project which was abandoned after the graduation of the two students (Bassan and La Rana 2017).

[90] For a discussion on the different techniques and prospects for developing detectors of very much higher sensitivity see especially section 4 and references therein in (Drever 1977).

[91] Weber's early resonant-mass antennae consisted of cylinders having the dimension of one to two meters which should vibrate at a specific resonance frequency when put in motion by a transit of a gravitational wave pulse. These devices were supposed to allow detection of a change in the cylinders' length by about $10^{-16}$ meters.

The seeds of this idea, as an alternative to Weber's antennae, can be found in a paper by the Soviet scientists Gertsenshtein and Pustovoit published at the beginning of the 1960s (Gertsenshtein and Pustovoit 1963), but the pioneer of this technique was Robert L. Forward, who got his PhD in physics from the University of Maryland in 1965, collaborating in the building and operation of Weber's bar antenna. Apparently the concept had been discussed within Weber's group around 1964 (Collins 2004, 265-266), but Forward was the first to build a small size interferometer in the late 1960s at Hughes Aircraft Company Research Laboratories in Malibu, and put in operation the first prototype detector in 1971 (Moss et al. 1971), improving it until 1978. He used a Michelson interferometer to look for changes in separation of masses about 3 meters apart. The sensitivity achieved was of order $10^{-13}$ cm in this baseline, which although inferior to that obtained with bar detectors designed for millisecond pulses, was encouraging in such a relatively small and simple system. He demonstrated that this idea could really work but did not obtain funds to move to a more sophisticated instrument.

---

[92] It is much more practicable to measure the distance between test bodies along one arm with respect to the distance between similar masses along a perpendicular arm, a layout particularly appropriate since the effect of a gravitational wave tends to cause the opposite sign of length change in the two arms. In a Michelson laser interferometer, a laser beam will be split into two identical beams by a partially reflecting mirror, with one beam reflected at 90-degrees from the first, but preserving the original frequency. Each beam travels down an arm of the interferometer and both are reflected back and merged into a single beam before arriving at the photodetector. As long as the arms do not change length while the beams are traveling, light waves will keep perfectly aligned canceling out in the recombined beam (totally destructive interference). Gravitational waves cause space to stretch in one direction and squeeze in a perpendicular direction simultaneously. For this reason, one arm of an interferometer will lengthen while the other one shrinks and constructive interference pattern will be observed in the photodetector. If one arm gets longer than the other, one laser beam will take longer to return back creating a phase difference between the two beams which will affect the interference pattern, showing that something happened to change the distance traveled by one or both laser beams. The interference pattern can be used to measure precisely how much change in length occurred and to extract information. The longer the arms of an interferometer, the smaller the measurements they can make. But this is an incredibly tiny effect, as gravitational waves, for example, can just change the length of a 4-km arm interferometer by $10^{-18}$ m, that is $1/1000^{th}$ the width of the classical proton radius. The trick is thus to create a longer light path which amplifies the gravitational-wave input to detectable amplitude: as long as the wave is passing, laser light in each arm bounces back and forth between the two mirrors hundreds of times before being recombined after such multiple passes. Nevertheless, detection of such small effect also implies that filtering out all possible sources of noise is one of the most challenging tasks for this investigative technique.

Simultaneously, Rainer Weiss from MIT had been actively exploring the idea of laser interferometry as a better chance of detecting gravitational waves since the end of the 1960s, starting a very detailed theoretical analysis of the ultimate sensitivity and of the noise sources of an interferometer. Interestingly, Weiss's project was not specifically connected with Weber's experiments: "My intent was never to check on Weber. The thing that excited me the most was the pulsars […]" (Collins 2004, 247).

After the failure of a first attempt in 1972, Weiss sent another funding application to the National Science Foundation (NSF) in August 1974, proposing the construction of a prototype interferometer with arms 9 meters in length.[93] Because of Kafka's deep involvement in the analysis and evaluation of Weber's experiment, he was asked to be one of the reviewers of Weiss' project. Controversially—and he later admitted this was an unfortunate breach of trust in his role as a reviewer—he circulated the proposal among the experimental groups in Munich (Collins 2004, 276–277).[94] At that particular moment in time, the group was actually investigating the possibility of designing an antenna that was to be kept at very low temperatures—near absolute zero—to reduce thermal noise, in parallel with other technical improvements to improve sensitivity (Biermann and Pinkau 1974, 105). For people who had worked on resonant detectors, the natural thing to do would have been to cool the detectors with liquid helium.[95] Billing, Kafka, Maischberger and Winkler were involved in long discussions on how to proceed further. They looked closely to cooled resonance detectors as well as to interferometers.[96] They were aware of the low temperature resonance experiments, as well as of Gerstenstein and Pustovoit's early suggestion of using interferometers as a means of detecting gravitational waves and of course knew about Forward's pioneering tabletop experiments (Gertsenshtein and Pustovoit 1963; Forward 1978). Walter Winkler well remembered the situation at the time:[97]

---

[93] Weiss' 1974 proposal, as suggested by Collins, was most probably "heavily based" on material appearing in the 1972 Quarterly Report of the Research Laboratory of Electronics at MIT (Weiss 1972). When mentioning attempts throughout the world aiming to confirm Weber's results with resonant gravitational antennae similar to those of Weber, Weiss specified, "A broadband antenna of the type proposed in this report would give independent confirmation of the existence of these events, as well as furnish new information about the pulse shapes. The discovery of the pulsars may have uncovered sources of gravitational radiation which have extremely well-known frequencies and angular positions." He calculated that the flux incident on the earth from the Crab Nebula pulsar was much smaller than the intensity of the events measured by Weber, but the detection of pulsar signals could be benefited by use of the new techniques he was proposing in his antenna design, which could "serve as a pulsar antenna." At that time, Weiss was also leading a balloon experiment to study the microwave background radiation, which is described in the first part of the Progress Report.

We had decided to stop the Weber-bar experiment, when it was clear that nothing could be found—despite of highly improved sensitivity compared to Weber's experiments. We were about discussing how to proceed, when Peter Kafka told us very reluctantly that he had got Rai Weiss' proposal to review, that he voted very positively to the American funding agencies. We had not yet made a decision about how to proceed experimentally and we were talking to each other every day. It would not have made much sense just to keep quiet. We therefore decided to inform Rai that we knew about his proposal, which indeed was for us a strong push into the direction of interferometry. Peter talked with Rai about our situation, and they understood each other quite well. In addition, Billing phoned Rai and asked him whether he would mind if we start to work on interferometers. Rai answered: No, why should I?

---

[94] See also Peter Kafka: Interview by Harry Collins available at http://sites.cardiff.ac.uk/harrycollins/webquote/, accessed 6/10/2018. Kafka felt somehow as an outsider in that field and "didn't understand much about the experimental possibilities" and so he "had to talk to the experimentalists anyhow", and then it was unavoidable that they discussed all these things in detail. According to Collins's interview with Robert L. Forward (also available at the same URL), Maischberger was involved as a reviewer, too, and he immediately thought of carrying out the interferometric experiment himself. It is interesting to observe how previous publications show that Maischberger was familiar with laser technology because he had worked at measurements with optical radars, radiating pulses of monochromatic laser light for atmospheric research, before beginning his work in gravitational waves. See for example (Bramanti and Maischberger 1972).

[95] On June 1, 1974, Biermann visited Pizzella and his group at their laboratories near Rome: "We showed Biermann the EXPLORER cryostat we were assembling and he was impressed and concluded that we were too much ahead, thus the German group would have done better to continue the search for GWs with a different technique, say with interferometers. It is worth to notice how much the premises of such a decision were wrong, since it took thirty years to the Rome group for reaching a sensitivity which was halfway with respect to the initial goal" (Pizzella 2016, 297).

[96] "Peter Kafka and myself thought at that time about the main limits for the sensitivity of large-scale interferometers for strain measurements. We worked out the standard quantum limit for such interferometers (to my knowledge nobody had done that before), when the sensitivity is limited equally by photon statistics and radiation pressure fluctuations. With reasonable assumptions—km size armlength, 10 kg mirror mass and gigawatt light-power—we ended up with strain sensitivities of $10^{-24}$ for short pulses. I described our considerations later on in more detail in my talk at the conference on Experimental Gravitation, in Pavia, September 17-20, 1976 (*Gravitazione Sperimentale*, Atti dei Convegni Lincei 34, Pavia, 1977)." Walter Winkler, personal communication to the authors, March 23, 2019.

[97] Walter Winkler, personal communication to the authors, March 18, 2019.

Actually, David Shoemaker, at that time working in Weiss' group, was later a member of the Munich/Garching project for 2 years.

The interferometric technique was more complex than the Weber bar. However, the Munich/Garching group was so enthusiastic about Weiss's plans that they immediately determined it would be possible to replicate the interferometric experiment using in-house resources. Certainly it was a courageous decision for researchers who had already given much effort in building Weber bars to start from zero again and explore a whole new technology.

A first indication of their intention of concretely moving towards a brand-new project can be found in the Annual Report of 1974: the development of more sensitive gravitational antennae was to use the principle of laser interferometry on which pioneering work started in Munich (Pinkau 1975, 103).[98] In March 1975, Kafka gave a talk at the International School of Cosmology and Gravitation in Erice, Sicily. He was very critical of Weber's results, showing that the current state of Weber bars (including the Munich-Frascati experiment), was a long way from achieving the optimal sensitivity required for detection. In mentioning the potential of extremely low-temperature detectors, Kafka pointed out that "laser interferometry with long 'free-mass antennas'" would be another avenue that seemed "worth exploring" (Kafka 1977, 239).

Kafka had been very positive in reviewing Weiss' proposal. However, much to their regret, again because of the failures of the American funding system to deal with research in controversial interdisciplinary fields, Weiss did not get the money from the National Science Foundation.[99] And so the original American project was delayed while the Munich group quickly moved forward with the new project.[100] Later, according to Kafka, the Americans themselves would use the Germans' success (and the fact that the project proposal had been inspired by Weiss' leaked proposal) to receive funding, and over the next decades, to an ever-increasing extent, they eventually took back control over the largest effort in gravitational wave detection experiments.[101] Walter Winkler recalled: "Rai Weiss stated in this respect: LIGO would not have been funded without the results from the Munich/Garching group."[102]

---

[98] See also (Kafka 1974c) and (Billing 1977).

[99] Weiss's opinion was that "The proposal to the N.S.F. was unfavorably reviewed at the time most likely because it was too big a step from acoustic gravitational wave detectors…" (Collins 2004, 275).

The Munich-Frascati coincidence experiment had achieved the highest sensitivities with room temperature bars. The work on laser interferometry that began between 1974 and 1975 was a possibility to improve the sensitivity by several orders of magnitude.[103] If cosmic gravitational waves could be detected, it was likely that they would reveal properties of their sources which could not be learned from electromagnetic, cosmic ray, or cosmic neutrino observations. By that time, all the different groups active in gravitational wave detections around the world (at Moscow State University, Yorktown Heights, New York, Rochester, Bell Labs, and Glasgow), failed to confirm Weber's detections, and thus it became general belief that his data were not to be ascribed to gravitational wave signals.[104]

---

[100] As Weiss later recalled, "They had been working on the bar detectors—the same method Weber used—and things had come to the end of the line with that. Virtually everybody in the world who had built the bar detector was seeing nothing […] And they were fetching around looking for the next step, and they were really turned on by this idea. So, they asked if there were people in my group who were working on it who would like to come to Germany. At the time, we hadn't gotten all the way, to where it was functioning. What happened, however, is that they started working on it. I mean, you can't stop people; you can't do that. And the Max Planck group in fact did most of the early development, because they had the money. I was always very jealous of that. They had the money, and they had a large group of very experienced professionals who had been working on Weber's kind of detector. And they went immediately into interferometers—this was about 1974, probably—and I couldn't go forward. So, I kept working more and more on the cosmic background radiation, because that's where I got money—having lost the military support." Rainer Weiss: Interview by Shirley K. Cohen, May 10, 2000. Transcript, California Institute of Technology Archives, Oral History Project, http://oralhistories.library.caltech.edu/183/, accessed 30/7/2019.

[101] Peter Kafka: Interview by Harry Collins available at http://sites.cardiff.ac.uk/harrycollins/webquote/, accessed 5/5/2019.

[102] Walter Winkler, personal communication to the authors, March 23, 2019.

[103] See the announcement about preliminary work on a laser antenna with improved sensitivity ("Vorarbeiten für Laser-Antenne mit wesentlich erhöhter Empfindlichkeit" [Preparatory work for laser antenna with significantly increased sensitivity]) in the section "Experimental Work" of the 1975 Annual Report (Kippenhahn and Pinkau 1976, 118). The group working on this project was formed by H. Billing, P. Kafka, K. Maischberger, L. Schnupp, W. Winkler.

[104] For an alternative view to Collins on this topic, see (Franklin 1994).

## 8. Parallel astronomical developments: A violent universe and the indirect observation of gravitational waves

In 1973, news about a new kind of signal, of unknown astrophysical origin, quickly spread through the scientific community (Klebesadel et al. 1973). The serendipitous discovery of gamma-ray bursts, intense fluxes of radiation clearly emitted in connection with catastrophic astrophysical events, not correlated either in time or position with any other known astrophysical phenomenon or object, generated a flurry of theoretical speculations. Several models were proposed, including shock waves in supernovae, stellar flares, collapsing neutron stars, comets captured by neutron stars, and many others.[105] The sources were likely to be associated with dwarf stars, neutron stars, or black holes. As with other transient astronomical observations, these could give clues to help understand the mechanism of the highest-energy processes in astrophysics—such as those associated with the final cataclysmic stages in stellar evolution. It was not until much later that this new message from the high-energy universe, not yet correlated to any optical event, would be connected to the simultaneous emission of gravitational radiation from cataclysmic events. At that point in time, it provided evidence of the possibility of radiation emitted from compact astrophysical objects, similarly to what was being hypothesized for gravitational waves.

Another astronomical discovery strongly encouraged the opening of a new hunting season. In January 1975, Russell A. Hulse and Joseph H. Taylor working with the radio telescope at Arecibo, Puerto Rico, announced the discovery of the first pulsar in a close binary system (Hulse and Taylor 1975). In opening up new possibilities for the study of relativistic gravity, it made it immediately clear that this system, now the most promising astrophysical source of gravitational waves, could provide the first test bed for strong field effects of general relativity.[106] Among the potential physical and astrophysical consequences, Einstein's theory predicts that over time such a system's orbital energy will be converted to gravitational radiation and the two stars will gradually spiral closer to one another as gravitational waves carry energy away.[107] The decrease of the orbital period (obtainable from the observed time

---

[105] Börner proposed as an explanation the bremsstrahlung (or braking radiation) of a beam of relativistic electrons hitting a region of high proton density (Anzer and Börner 1975).

[106] Since 1962, it had been suggested that a double system with one white-dwarf component could radiate enough gravitational power to be detectable and even become a test for the existence of gravitational waves (Kraft et al. 1962).

[107] See previously mentioned article by Dyson written in 1963, when pulsars—and consequently neutron stars—

variation of the pulsar period) would thus constitute "a test for the existence of gravitational radiation" (Wagoner 1975).[108] Such frontier astronomical phenomena were pushing Einstein's solar-system-tested general theory of relativity to explore a much wider environment. As a result, it was expected at the time that gravitational radiation might become a powerful tool for observational astronomy.[109] But its impact went well beyond. With the discovery of pulsars, quasars, and galactic X-ray sources, and the coincident expansion in the search for gravitational waves, relativistic gravity—which had always played a central role in cosmology—was assuming an important place also in the astrophysics of localized objects.

In March 1976, while observations with the Weber-type resonant antennae ended, a 3-meter interferometer was being built by the Munich group.[110] Their Weber-type coincidence experiment had been run between July 1973 and February 1976. Reporting the total result up to the dismantling of the detectors, they compared these with "future aims of gravitational pulse astronomy," the most interesting of which was expected to be the observation of stellar collapse:

While the observation of 'weak' radiation, e.g. from close binaries, will be helpful for a confirmation of Einstein's theory (and a check of the approximation methods to solve its

---

had not yet been discovered, suggesting that a close binary system of neutron stars could emit a powerful burst of gravitational waves after coalescing of the two components (Dyson 1963).

[108] At the time, Ehlers stressed that "the state of the theory of gravitational radiation itself was by no means satisfactory; relativity could not properly be tested against the observations until relativists sorted out the theory" (Directors of AEI 2008). See also (Ehlers 1974). Hulse and Taylor were awarded the 1993 Nobel Prize in Physics for the discovery of the binary pulsar.

[109] For a review of the history of the discovery of the first binary pulsar and a description of its immediate impact and its longer-term effect on theoretical and experimental studies of relativistic gravity, see (Damour 2015). Kennefick has given a full account of how the measurement of orbital decay in the binary pulsar PSR 1913+16 interacted with an ongoing debate amongst theorists about whether the quadrupole formula, expressing the rate of emission of gravitational wave energy by a system of accelerating masses, could give a reasonable approximation of the source strength of possible astrophysical sources of gravitational waves, especially binary stars. References to previous work are in (Kennefick 2017). Kennefick has remarked how by the mid 1970s most theorists accepted that binary star systems did generate gravitational waves, but still some experts doubted whether the quadrupole formula could be correctly applied to them. Ehlers and other theorists were in fact among the skeptics objecting that "a formula for the energy loss due to gravitational radiation of bound systems such as binaries had not yet been derived either exactly or by means of a consistent approximation method within general relativity, a view which contradicts some widely accepted claims in the literature […] derivations presented so far either contain inconsistencies or are incomplete" (Ehlers et al. 1976, L77).

[110] The working group comprised Billing, Kafka, Maischberger, Schnupp, and Winkler (Kippenhahn and Pinkau 1977, 142). The new project was presented at the international meeting on experimental gravitation held in Pavia in 1976 (Winkler 1977). By 1977, the section reporting research on "Gravitationstheorie" had expanded into several diverse research lines and the members of the gravitational wave group were dividing their attention between studying various different technical problems (Kippenhahn and Pinkau 1978).

equations), the 'strong' radiation from final collapse will contain fascinating additional information about the behavior of matter at extreme densities, unobtainable in any other way (Kafka and Schnupp 1978, 103).

The authors had compared the expected strengths and rates of gravitational wave signals from core collapse in supernovae with sensitivities of the then current detectors, whose performance—according to their evaluation—could not be improved in order to be able "to detect events at the rate of several per year or greater." Consequently, they had decided "not to continue with (low-temperature/high-quality) Weber-type experiments, but rather with a Weiss-Forward type experiment," i.e., a laser-lighted Michelson interferometer:

It is hoped that the many new antennae, being developed now and during the next decade, will be able to detect a few gravitational collapse events per year, and thus provide most valuable information on extreme states of matter and final stages of stellar evolution which will otherwise remain hidden (Kafka and Schnupp 1978, 103).

In December 1978, the Ninth Texas Symposium on Relativistic Astrophysics, which had become the principal international meeting where relativists and astrophysicists met and discussed recent research, was held in Munich (Ehlers et al. 1980).[111] For the first time in the history of these series of meetings, a Texas Symposium was held not just outside Texas but also outside the continental United States.

An outline of research fields studied at the Institute for Astrophysics at that time included the old classic 'battle horses' such as atomic and molecular physics, solar physics, comets, star formation, and end phases of star development, as well as the new relativistic sector: gravitational theory and relativistic astrophysics, quasars, supernovae and collapse, gravitational waves (Kippenhahn and Lüst 1977).[112] Since the 1960s, much effort had been made concerning the modelling of collapsing stars and the evaluation of their gravitational radiation emission, as they had been considered the best candidates for the production of frequent and intense gravitational pulses. These topics were discussed at the meeting, with the program also including microwave-background radiation related to the dense, hot initial stage of the universe, a survey of the gamma-ray sources, as well as various theoretical and observational aspects of X-ray astronomy linked to late stages of stellar evolution, particularly in binary star systems. One of the workshops organized within the different sections was dedicated to gravitational radiation. In a wide overview of the then current status of relativistic astrophysics and of the latest scientific developments discussed during the conference, Börner and Kafka remarked that the previous 15 years had also brought two Nobel Prizes, which had never before been received in astronomy.[113] They duly commented how "Relativistic astrophysics has no sharp boundaries. Its best definition is probably still the program of the Texas symposia" (Börner and Kafka 1980, 181).[114]

---

[111] In Reimar Lüst's papers see a proposal for the conference written by Jürgen Ehlers to Jürgen Buntfuss, German Research Foundation (DFG), on January 14, 1976 (AMPG, III. Abt., Rep. 145, Folder 230 "Korrespondenz Klaus Pinkau", Fol. 253). See also (Börner and Kafka 1980). The event received financial support from the Max Planck Society (through its President, Reimar Lüst), the German Research Foundation, and the Institute for Astrophysics directed by Kippenhahn. Support was also provided by the Bavarian government and the city of Munich.

[112] See also the previously mentioned annual reports from the 1970s/early 1980s.

The Texas Symposium held in Munich was also the occasion of the first public announcement of the experimental evidence for the reality of gravitational radiation damping in the binary pulsar discovered by Hulse and Taylor, which was published shortly afterwards (Taylor et al. 1979).

## 9. Scaling up interferometry: an itinerant gravitational wave group in the 1980s

Ludwig Biermann officially retired in March 1975 but continued to be active at the institute. In promoting the gravitational wave experiment, he had added a last fruitful item to his rich long-lasting legacy.[115] Starting around 1974, the Munich/Garching group had built prototypes of interferometric gravitational wave detectors to find solutions for occurring problems and develop improvements for existing techniques and instruments—a new field of research in many respects.[116] Since the previous years, discussions on continuing research on the gravitational wave experiment with laser interferometry were ongoing, also in view of Heinz Billing's retirement in 1982.[117] The committee on the future of Billing's group expressed the opinion that the prototype 3-meter gravitational wave antenna was a project of fundamental importance and proposed continuing with the preliminary phase, during which the small interferometer would be tested in view of the more ambitious project for a 30-meter antenna (Rüdiger et al. 1987). The group worked hard to reduce several unwanted effects and developed innovative technologies that then had an impact on future gravitational interferometers.[118] They eliminated the excess noise reaching the shot-noise level in 1982 (the lower limit of detection set by the quantum noise effect originated from the discrete nature of photons and electrons), which meant they had "found out *all relevant noise sources*, understood them well enough, found means and ways to reduce them sufficiently for the shot-noise level at that time and compatible with the rest of the setup."[119]

---

[113] They were referring to the Nobel Prize in Physics 1974, awarded to Martin Ryle and Antony Hewish, for their research in radio astronomy and the discovery of pulsars. Jocelyn Bell, who had in fact been the first to observe and analyze the pulsating radio signal, was excluded from the prize. The second Nobel Prize for an astronomical discovery was awarded to Arno Penzias and Robert Wilson in 1978 for the discovery of the cosmic microwave background radiation.

[114] See also Kafka's two review articles (Kafka 1979a, 1979b).

[115] Biermann became emeritus on March 31, 1975. In the Annual Report, signed by his successor Kippenhahn, his instrumental role during almost 30 years at the Institute for Astrophysics in opening and promoting new research fields—ultimately leading to the foundation of new Max Planck institutes—was emphasized also recalling the impulse given with the establishment of the department for relativity theory and the promotion of the Munich gravitational experiment (Kippenhahn and Pinkau 1976, 112).

[116] Winkler has remarked, "For the first time the optical components have been individually suspended. A properly shaped laser beam is injected into the interferometer, the beams in the two arms have to be properly adjusted

---

and are eventually brought to interference. Servo-systems have been developed and implemented to operate the interferometer at optimal interference—the dark output—and keep it there. The first interferometer was a rigid block, containing beam-splitter, two mirrors and Pockels cells to adjust the lightpath in the arms for destructive interference at the output port. With this rigid interferometer 4 noise sources could be identified: 1. Laser intensity noise, 2. Laser frequency noise, 3. Laser motions of the laser beam, 4. The so-called parasitic interferometer – a reflex superimposing with the original laser beam. In October 1978, 3-m arm-length was started and delay lines were added to the arms to increase the optical pathlength, as proposed by Rainer Weiss." Walter Winkler, personal communication to the authors, April 27, 2019. He also emphasized that the operation and results obtained with the prototypes are described in his talk at the Ninth Texas Conference held in Munich in 1978, and in (Billing et al. 1979b). Described in detail are local and global control of mirror position and orientation, optical feedback for path difference stabilization, frequency stabilization of the laser light, influence of scattered light and resulting requirements for the setup. Specifications for the delay-line mirrors had been set up. Fluctuations in beam-geometry (lateral displacement, orientation, beam-width) and resulting spurious signals had been investigated for the first time. They made approaches to stabilize the beam in this respect by introducing a beam-symmetrizer, which led eventually to their invention of a mode-selector to stabilize the geometry of a laser beam.

[117] The fate of Billing's group, still named "Numerische Rechenmaschinen," was discussed starting from March 1977 (CPTS meeting minutes of 08.03.1977, 22.06.1977, 01.02.1978, AMPG, II. Abt., Rep. 62, No. 1780, 1781, 1783).

[118] Some construction details of the prototype of 3-meter arm length used for early studies of noise and other disturbances, such as laser frequency instabilities, might restrict signal perceptibility (Maischberger et al. 1979; Billing et al. 1979b). The new project was presented in the internal report (Billing et al. 1979a). See also Heinz Billing et al., "The Present State of the Munich Gravitational Wave Experiment," in (Schmutzer 1983, 401). In the Annual Report for 1980, the 30-m device was presented as an intermediate stage to a 300-m arm-length interferometer (Kippenhahn and Fink 1981, 229). Ongoing research was presented in internal reports before being published (Rüdiger et al. 1980; Schilling et al. 1980, 1981; Maischberger et al. 1981).

[119] Walter Winkler, personal communication to the authors, March 23, 2019. An example in this sense, added Winkler, was the suspension of the mirrors: "When we started around 1974/1975 to suspend the optical components like mirrors or the beam-splitter as pendulums (in order to isolate them from mechanical noise), it was immediately clear that further isolation stages have been necessary to avoid the excitation of the different degrees of freedom well above the thermal excitation. Therefore we used right from the beginning several mass-spring components in series in addition to the pendulum mode. Later on we have used triple pendula. In the beginning we damped by so called local damping via coil/magnet systems several degrees of freedom, also that

This first 3-meter prototype "was the best in the world for many years" (Collins 2004, 277).

With Billing's retirement, the heroic era of gravitational wave experiments at the Institute for Astrophysics was coming to an end and at the same time, the development of laser interferometers was changing globally the scale of gravitational wave experiments.[120] Moreover, the challenge to detect gravitational waves was creating a new chapter in the field of quantum electronics. By October 1980, a decision had been taken to move the gravitational wave experiment group to the Max Planck Institute of Quantum Optics, which was founded on 1 January 1981.[121] In August of that year, an international meeting on quantum optics and

---

of the suspension point of the lowest pendulum. In addition we used coil/magnet systems and Pockels cells in the light path to keep the interferometer at its point of operation. Somewhat later we invented the shadow-meter: a small plate interrupting a part of a light-beam, which falls onto a photo-diode. A motion of the plate changes the amount of light, falling onto the photodiode. Thus the photocurrent changes correspondingly to the relative motion. This shadow-meter was subsequently used by the other groups. As usual, the mirrors themselves had been fixed on some kind of mirror holders. When looking at the interferometer signal, we found huge noise contributions coming from the resonances between mirrors and their holders. Whatever we tried out – nothing really helped. One day Karl Maischberger said to me: why not suspend the bare mirrors in a wire sling, and thus avoid these ugly resonances? We did so, and immediately the noise level was much better. An ingenious idea, which nobody had thought of before. Then we had to invent means to adjust the mirrors properly and keep them there. Later on this arrangement was improved by the Glasgow group, suspending the fused silica mirrors on fused silica wires, and fix them together by silicate bonding, thus making up a monolithic component with high mechanical Q. At that time, we had also solved all the relevant problems like stabilization of the laser beam in frequency (in the end relative to the light-path inside the interferometer) and geometry, servo systems, scattered light contributions, vacuum requirements, data acquisition etc. Otherwise we would not have got the shot-noise level as set by the laser-power at that time!"

[120] During the meeting of the CPT section on October 29, 1980, it was reported that the committee on the future of Billing's *Rechengruppe* was of the opinion that "The gravitational wave experiment is of fundamental importance and therefore recommends that the preliminary experiment be continued. In the event that it proves to be promising at the time of Mr Billing's retirement, the Commission asks the President to ensure that the main experiment is continued inside or outside the Max Planck Society." It was also noted that "This assessment was confirmed by foreign experts. Scientists from Caltech and MIT had advised the president in talks to continue the experiment. Based on the positive statements, the project should be continued in the Max Planck Society. Mr. Walther agreed to take the group into the Institute of Quantum Optics. Finally, the Chairman noted that the Group currently holds the top international position with its work" [translations by the authors]. CPTS meeting minutes of 29.10.1980, AMPG, II. Abt., Rep. 62, No. 1791. On March 10, 1980 Winkler gave a talk at the advisory board meeting at the Institute for Astrophysics about the status of the gravitational wave experiment. They had just reached the shot-noise limit of 50 mW laser-power and had found the fundamental importance of scattered light. Reimar Lüst, at that time president of the Max-Planck Society, was present and "was obviously ready to support the research after Billing's retirement in 1982." Walter Winkler, personal communication to the authors, April 4, 2019.

[121] In fact, the roots of the Max Planck Institute of Quantum Optics dated back to the establishment on January 1, 1976 of a Laser Research Group set up at the Max Planck Institute for Plasma Physics (IPP) as a result of an agreement between the German Federal Ministry for Research and Technology, as it was called at the time, and the Max Planck Society. The aim of such a group was to work on the development of high-power lasers and their application to plasma physics, chemistry, spectroscopy, and other fields. This issue was discussed at the meetings of the Max Planck Society's 'Senatsausschuss für Forschungspolitik und Forschungsplanung' (Senate committee on research policy and research planning) in 1975 (see copies of the minutes in AMPG, III. Abt., Rep. 68 A, No. 151). The committee discussing the future of this group and its transformation into the Institute of Quantum Optics with Karl-Ludwig Kompa, Herbert Walther, and Siegbert Witkowski as Directors, was formed on June 14, 1978

experimental gravity was organized by the new institute and promoted by the NATO Advanced Study Institute on Quantum Optics and Experimental General Relativity. The meeting aimed at establishing links between physicists working in fields traditionally separated as quantum optics, experimental gravitation, and the quantum theory of measurement. Efforts to develop gravitational wave detectors already underway in several laboratories around the world—and their quantum mechanical nature coming into play because of the weakness of the signals they were attempting to measure—were unifying these previously far removed areas. Joint discussions during the meeting offered the opportunity to "close the gap" (Meystre and Scully 1983).

In May 1982, when the gravitational wave group became officially part of the Max Planck Institute of Quantum Optics in Garching, construction of a new prototype interferometer, which would have a 30-meter path, had already started. It was completed in mid-1983, but improvements continued to be made over the years (Max-Planck-Gesellschaft zur Förderung der Wissenschaften 1983, 701).[122] The scaling from the 3-m to the 30-m device worked exactly as expected, giving the confidence for building larger detectors. Weiss himself expressed the valuable efforts made by the group: "So the Max Planck group actually did most of the very early interesting development. They came up with a lot of what I would call the practical ideas to make this thing better and better."[123] He further remarked: "The Germans have found and solved problems which I had not even thought of!"[124]

---

and during the CPT Section meeting of May 5, 1979, the final formal decision was unanimously taken (CPTS meeting minutes of 14.06.1978, 30.01.1979, 09.05.1979, AMPG, II. Abt., Rep. 62, No. 1784, 1786, 1787). In 1981, the research group was given separate status as the Institute of Quantum Optics and the Research Group on Gravitational Waves became involved with the development of laser interferometers. The group at IPP initially had 46 members and quickly grew to 105, so that the space made available by IPP, including additional barracks, soon became too small. In 1986, when the institute moved to a dedicated new building, there were 184 staff members. See preface and Section 3.2.10, entitled "Messung von Gravitationswellen – eine Revolution in der Astronomie?" in (Max-Planck-Gesellschaft 1986).

[122] A description of the laser interferometric project related to that stage of activities can be found in (Billing et al. 1983) and (Rüdiger et al. 1983). See also the later internal reports: (Schilling et al. 1984; Shoemaker et al. 1985; Schilling et al. 1984; Shoemaker et al. 1985, 1988). David Shoemaker, who joined the Garching group developing the laser interferometer, had worked with Rainer Weiss at MIT on the early 1.5-meter prototype (Livas et al. 1986). For a short history of the institute, see preface in (Max-Planck-Gesellschaft 1986).

[123] Reiner Weiss: Interview by Shirley K. Cohen, May 10, 2000. Transcript, California Institute of Technology Archives, Oral History Project, http://oralhistories.library.caltech.edu/183/, accessed 19/1/2019.

[124] "When I was member of STAC [Scientific and Technical Advisory Committee] for Virgo—remembers Winkler—I recommended the people to build a prototype interferometer. The answer at that time was: We do not need one. We simulate everything on the computer. But: You will not find all relevant problems and a solution for them just by working at a computer! You never will think of the influence of scattered light (we have found it

In the meantime, the observation between 1974 and 1981 of the first binary pulsar system discovered in 1974 had clearly demonstrated that the orbit was slowly shrinking, following the curve predicted by general relativity for the loss of energy and momentum due to gravitational wave emission (Taylor and Weisberg 1982). This progress, together with tremendous advances in experimental tests of relativity, contributed to stringently constrain or even rule out alternative theories, increasing the confidence of gravitation theorists that general relativity was the correct classical theory of gravity. Despite the fact that detection of gravitational radiation remained to be demonstrated, this was the first indirect proof of the existence of gravitational waves, providing strong support for decisions to start launch more ambitious projects.

**10. Munich's initiative (and failure) to build a km-scale interferometer**

A new group in Italy, led by Adalberto Giazotto, was working at the University of Pisa from 1982 onwards on a seismic noise attenuation system for very low frequencies—which were supposed to be emitted by several pulsars—in view of a future gravitational-wave interferometer.[125] The Italians began to discuss a joint project with the French group headed by Alain Brillet during the Fourth Marcel Grossmann Meeting on General Relativity held in Rome in 1985 (La Rana and Milano 2017, 191). The French had started in the early 1980s a prototype project in Orsay, near Paris, investigating lasers and the interferometric approach with the goal to operate a 5-10 m prototype within 5 or 6 years (Brillet 1984). Their complementary expertise led to an Italian-French collaboration and to the definition of a project for an interferometric antenna led by Brillet and Giazotto in 1989 (Bradaschia et al. 1990).

The group at Glasgow University, too, had moved towards the development of techniques for the detection of gravitational radiation using optical interferometry since 1975. Like in

---

in connection with a reflex from an antireflectively coated area superposing with the main beam. I calculated the problem through and found that a few photons are in principle sufficient to produce spurious signals). Another example: geometrical motions of the laser beam relative to the interferometer cause signals via slight asymmetries. Nobody had thought of that before. We invented the mode-cleaner. There are many of those experiences we had, and which nobody had thought of before." Walter Winkler, personal communication to the authors, March 23, 2019.

[125] See Adalberto Giazotto: Interview by L. Bonolis, Pisa, December 18, 2006. Transcript in (Bemporad and Bonolis 2012) Published online at http://static.sif.it/SIF/resources/public/files/uomini-quarks/giazotto.pdf, accessed 18/1/2019.

Garching, the strategy had been based on the hope that, once sophisticated prototypes of modest length had been operated successfully, the sensitivity to gravity waves could be improved fairly rapidly by scaling up the length of the arms, without making major changes in the instrumentation by which the length difference was monitored. For this reason, all effort focused on developing the monitoring instrumentation on prototype detectors of small arm length. At Glasgow they had built and were developing a system with an arm length of 10 m. A special attention was paid to identifying all noise sources, understanding them thoroughly, and devising ways to remove them which would work not just on the prototypes, but also on much larger future detectors (Robertson et al. 1982; Drever et al. 1983; Hough et al. 1983, 1984, 1986).[126]

In 1987, the detection of neutrinos from the supernova 1987A appeared to be a failed opportunity for gravitational wave detections.[127] As Kafka commented: "None of the more sensitive of the presently-existing GW antennae was working on February 23$^{rd}$ 1987 when the supernova in the Large Magellanic Cloud went off—not to mention the more sensitive antennae planned for the near future. Otherwise the birth of GW Astronomy might have been registered" (Kafka 1989, 55).[128]

---

[126] In 1979 Ronald Drever took up a part-time appointment to Caltech, and full-time later, in 1983, leaving James Hough as the Glasgow leader. At Caltech Drever started a project which was eventually funded.

[127] This was the closest observed supernova since the seventeenth century. It became the first instance of detection of neutrinos in such an event, reported by four underground laboratories around the world, all within 24 hours of the visual sighting. See (Woosley and Phillips 1988) and references therein. As Wheeler had already stressed during a conference on underground science in 1982 (Wheeler and Wheeler 1983, abstract): "At least one kind of supernova is expected to emit a large flux of neutrinos and gravitational radiation because of the collapse of a core to form a neutron star […] The corresponding neutrino bursts can be detected via Cerenkov events in the same water used in proton decay experiments. Dedicated equipment is under construction to detect the gravitational radiation. Events throughout the Galaxy could be detectable, but are expected only at intervals exceeding a decade. Nevertheless, the next event could come tomorrow, so every attempt should be made to make the monitoring for such events routine."

[128] Kafka's contribution was included in the proceedings of a NATO Advanced Research Workshop that was held from 6 to 9 July 1987 in Cardiff, representing a snapshot of the state of the gravitational wave community's thinking and understanding in the summer 1987. Bernard Schutz, who had been the promoter of the workshop, following discussion he had in 1985 and 1986 with many of the principal members of the various groups building prototype laser-interferometric detectors, emphasized in the preface that even if most of the effort had been concentrated on the detector system, proposals being planned by the different groups would have to address also questions related to computer hardware required to sift through data coming in at rates of several gigabytes per day and what software would be required for this task. Moreover, given that every group had accepted that "a worldwide network of detectors operating in coincidence with one another was required in order to provide both convincing evidence of detections of gravitational waves and sufficient information to determine the amplitude and direction of the waves that had been detected, what sort of problems would the necessary data exchanges raise?" Schutz further remarked in the last lines of the preface: "None of us knows when the *first* gravitational wave will be observed in our detectors, but as the book shows, we are already looking beyond that momentous

But the gravitational-wave community was laying the premise to get the proper sensitivity requirements for the future observation of such catastrophic events, since the two main candidate sources for detection of gravitational radiation were supernovae and coalescing of close binary systems composed of highly condensed partners, i.e. neutron stars or—even more efficient—black holes. In Glasgow they were considering the possibility of building a larger detector of arm length approximately 1 km (Hough et al. 1984), and in Garching, after encouraging progress with the 30-meter prototype, the group was stepping up efforts in order to prepare for a big jump in size: a full-sized 3-km arm-length interferometer.[129] Preliminary investigations for this ambitious project ("Voruntersuchungen für den Bau eines großen Laserinterferometers zur Messung von Gravitationswellen") led by Gerd Leuchs at MPI for Quantum Optics, were financed by the German Federal Ministry for Research and Technology (BMFT) during the period 1987-1989.[130]

Both groups had gained considerable experience in the design and operation of prototype versions of interferometric detectors since the early 1970s. The experimental group at Glasgow had benefited from collaboration with a theoretical group led by Bernard F. Schutz at the University of Wales at Cardiff, interested in analysis of signals from such detectors, and in Garching, right from the start, the experimental group had been in close contact with colleagues at the Max Planck Institute for Astrophysics—where the gravitational wave project was born—especially with the Department of General Relativity led by Jürgen Ehlers.

---

event to the establishment of *gravitational wave astronomy*, the regular detection and identification of gravitational waves from a great variety of different sources scattered throughout the universe (Schutz 1989, preface). See also Schutz' discussion about possible strategies for maximizing coincidences between detectors in the U.S. and Europe in (Schutz and Tinto 1987).

[129] The concept of the large antenna was described in (Maischberger et al. 1985). Albrecht Rüdinger presented the project at the Fourth Marcel Grossmann Meeting on General Relativity (Winkler et al. 1986). Plans for the large detector were described also in (Rüdiger et al. 1987; Maischberger et al. 1988). A definition phase, with an expected duration of between 1 and 2 years, was beginning. During this period, various questions would be clarified, including different technical issues as well as the choice of the site and a reliable estimate of the main cost items (Winkler et al. 1985). The report MPQ 96 was the precursor of a later report, the updated study MPQ 129, organized in three main parts preceded by a summary providing information about the content (Leuchs et al. 1987a). The English translation of these chapter summaries was included in a further report, MPQ 131 (Leuchs et al. 1987b). The authors thanked Peter Kafka for writing the first introductory part reviewing the physics and astrophysics of gravitational waves in the context of the proposed big antennae, which was also published in two articles in *Die Naturwissenschaften* discussing the general properties of the waves and the planned antenna sensitivity, in connection with the expected sources (Kafka 1986a, 1986b).

[130] As part of a series of high-energy physics projects, DESY (Deutsches Electronen-Synchrotron), the German national research center operating particle accelerators in Hamburg, was supporting the execution of the project as lead partner on behalf of BMFT (AMPG, II. Dpt., Rep. 66, No. 3122, 3853, 3868).

But during 1988 it became clear that the British proposal for a 1-km antenna would not be financed by the Science and Engineering Research Council (SERC).[131] Because of serious financial problems, the funding of such an expensive enterprise was in fierce competition with projects put forward by the astronomy/astrophysics community.[132] But still around 1989, in Germany, "the idea of building a large interferometer crystallized to be physically highly interesting, technically viable […] financially within range of becoming reality".[133] It could be envisaged a cost sharing between the regular budget of the Max Planck Society, a grant from the BMFT and a support by the state.[134] As the British group had given up plans for its own large project, a fourth partner was now in sight. And so, the Garching project for a 3-kilometer interferometric gravitational wave detector resurfaced in 1989 as a joint German-British proposal (Hough et al. 1989), strongly encouraged by the two funding bodies BMFT and SERC. In Appendix A of the proposal the two groups presented the results of a 100-hour period of coincident observation using the two prototypes at Garching and Glasgow (the 30-m and the 10-m arm length), which had been solicited by BMFT and SERC to show that such detector could be operated in the production fashion by the two teams working together. It was the first time that two detectors had been run continuously in a data-taking mode, demonstrating the potential for long-term operation of laser interferometric detectors. But the beginning of a long period of economic recession in U.K., starting during 1990 and going on until spring 1993, would strongly influence the destiny of the German-British project. By mid 1991 it would become clear that the U.K. would not be able to contribute with funds at least for a couple of years. As we see in the following pages, this lack of British commitment had the effect of making the Max Planck initiative vulnerable during these crucial years: while the proportion of SERC financial contribution was relatively small (20 million as opposed to 100 million of the BMFT), the Ministry had been insisting since the 1980s that any large scientific project should be done as an international collaboration; this financial impasse could be mobilized by skeptics of the gravitational waves enterprise in Germany to slow down its advance, diminishing the need for direct scientific confrontation.[135]

---

[131] For a general overview of ongoing projects and state of art of the field at the time see (Hough et al. 1987).

[132] At the end of the 1980s, many conventional astronomers were still very suspicious and did not consider gravitational waves as something worth funding, a circumstance which influenced such decisions. The cosmic-ray physicist Alan Watson was one exception, and his support proved crucial in the 1990s, when the GEO collaboration joined LIGO. Bernard Schutz: Interview by Adele La Rana, CERN, August 28, 2017, and personal communication to the authors, 24 November 2019. This was also the case in Germany, and was reflected in the Denkschrift (white paper) which should guide astronomical investments in the coming decades. Heinrich Völk's recollections of the time he was editing the *Denkschrift Astronomie* (Völk et al. 1987) in collaboration with his good friend Peter Biermann, son of Ludwig Biermann, are very explicit about this point: "We also tried a little bit to get things which we found interesting, like Gravitational Wave astronomy. We had to fight hard to get that in! You cannot imagine what kind of… opposition you encounter in such a case. Anyway we put gravitational waves as one of the—not the most expensive—but the most important projects in this. I'm still proud of it, that we did that at the time!" Heinrich Völk: Interview by Luisa Bonolis and Juan-Andrés Leon, Heidelberg, October 9-10, 2017. About US astronomers and astrophysicists animosity towards LIGO and how they felt that the project was competing for their resources see (Collins 2004, 500-504).

[133] Hermann Schunck: Written interview by Adele La Rana, May 14, 2019.

[134] Funds for the preliminary investigations for the construction of a large laser interferometer were provided by BMFT and searches for a suitable site went on during the second half of the 1980s. See documents on the financing of a grant ("Voruntersuchungen für den Bau eines großen Laserinterferometers zur Messung von Gravitationswellen") starting in November 1987 and ending in December 1990 and for a second tranche covering the period from 1.1.1990 to 31.12.1992 (AMPG, II Abt., Rep. 66, No. 3853, 3868 and Rep. 68, No. 65). Collaboration with other European groups, in particular with the Glasgow team, is also mentioned in the proposal by the MPI for Quantum Optics to BMFT for the period 1990-1992, related to a requested sum of 4.184.500,00 DM (about 2 million Euros of today). See letter sent on December 13, 1989, from the Max Planck General Administration to DESY (addressed to Dr. Prünster), the research institution acting as project-executing agency on behalf of BMFT, in which it was specified: "We [MPS] are unable to put at disposal of the Institute [of Quantum Optics] additional funds for this project. We support the application and would be very grateful to you for promoting this project." (AMPG, II Abt., Rep. 66, No. 3853).

The gravitational wave communities around the world in the early 1990s navigated a political-rhetorical minefield as they simultaneously presented their cases to the respective funding bodies: The argument was made by all of them for the need of several detectors around the world, which gave support to each other's projects; but at the same time this was difficult to reconcile diplomatically with ambitions of global leadership. The Germans during these years were at a particular disadvantage to the French-Italians, while at the same time aiming to negotiate as equals with them based on the premise of having their own detector.

As stated in the preface of the joint proposal (Hough et al. 1989), it was expected that "all the long baseline detectors to be built [the LIGO project and the Italian/French Virgo project] will operate as part of a coordinated worldwide network." At that time, future prospects for the realization of a big interferometer still looked excellent. From 1990, the gravitational wave project at the Max Planck Institute of Quantum Optics in Garching was led by Karsten Danzmann, who had come back from Stanford University, where he had moved in 1982 after his PhD at the Technical University in Hannover.[136] Gerd Leuchs, who had led the Garching

---

[135] See folder on the gravitational-wave experiment in the Archives of the Max Planck Society in Munich: Akten der Registratur und des Archivs der Max-Planck-Gesellschaft, MPI für Quantenoptik, Gravitationswellenexperiment III, 1991-1997, Aktenzeichen 18140907, Barcode 233163 (from now on ARMPG), Fol. 373-380.

[136] After working in plasma physics and astrophysics, Danzmann had turned to laser spectroscopy. Immediately after having listened to Danzmann's talk at a conference on laser spectroscopy in the U.S., Herbert Walther,

group from 1985 to 1989, moved to work into industry and later became Director at the Max Planck Institute for the Science of Light.

In September 1991, the German-British project, now named GEO, was presented at a meeting organized in Bad Honnef by Ehlers and Gerhard Schäfer as a 3-km arm-length interferometer to be built near Hannover, in the German state of Lower Saxony (Danzmann et al. 1992).[137] In March 1992, the French CNRS, the Italian INFN and the German MPG signed an "Expression of common interest", an agreement to promote "an effective collaboration between the European teams in view of building and operating two antennas in Europe: the French-Italian project VIRGO and the German-British project GEO."[138] In Summer 1992, a 3-km GEO interferometer was still part of a list of the detectors at that scale being planned in the world: the French-Italian 3-km Virgo (comprising 9 groups from both countries) to be built near Pisa, the American 4-km LIGO (Laser Interferometer Gravitational-Wave Observatory) project (approved in fall 1991) with scientists at MIT and Caltech, and a more recent Australian collaboration proposing a 3-km detector near Perth (AIGO, not yet approved at the time). Plans for these full-scale astrophysical observatories required the evolving of the prototypes from laboratory setups used to test new optical measurement techniques into stable astrophysical instruments. They were meant not to be "in competition with each other", on the contrary, each of them was considered crucially dependent on the others, since it had been evaluated that "to fully unravel the information contained in the signals with respect to the source direction, time structure and polarization" required "a world-wide network of four detectors." The hope was that the network could be in place by the end of the 1990s, and that at the beginning of the next millennium they might be able "to mark the beginning of the age of Gravitational Astronomy" (Danzmann 1993, 19).

---

Director of the Max Planck Institute of Quantum Optics, told him "Mr. Danzmann, you will come to Munich and work on gravitational waves!" Interview with Karsten Danzmann, March 29, 2018, Deutsche Physikalische Gesellschaft e. V., Stern-Gerlach-Medaille 2018, available at https://www.youtube.com/watch?v=tNTB74bFGuc, accessed 23/2/2020.

[137] See also (Lück and the Geo600 Team 1997; Völter 2016).

[138] ARMPG, Fol. 322-323. And while Heinz Riesenhuber, the Minister of Scientific Research, was favoring a pan-European solution, the proposed German (GEO) project "seemed not feasible in the current form", as from a memorandum dated October 19, 1992, prepared for a meeting to be held in Paris to discuss the European gravitational-wave project EUROGRAV" with the participation of France, Italy, Germany, Great Britain (and Niedersachsen), and which should "without any doubt" preview the building of two detectors (ARMPG, Fol. 391-392). See in the same folder several documents testifying high-level interactions aiming at consolidating such agreements.

At a time of enormous expansion of interest in, and importance of, Einstein's theory of gravitation, the next major step should be the construction of a number of long-baseline detectors around the world. An array of detectors of this type was expected "to allow the observation of gravitational waves from a range of astrophysical sources, leading to improved insight in many areas including stellar collapse, binary coalescence and the expansion of the Universe."[139]

However, in spite of contacts between the European groups going on during the second half of the 1980s, the Italian-French collaboration and the British-German venture had not merged into a real joint pan-European effort, a European network of gravitational-wave telescopes that might have followed and matched the successful example of effective cooperation in the CERN enterprise.[140]

By 1992, the existence of gravitational waves had been demonstrated through the motion of the double neutron star system PSR 1913+16, in which one of the stars is a pulsar emitting electromagnetic pulses at radio frequencies at precise, regular intervals as it rotates. Arrival-time measurements of the radio signals running since 1974, showed an orbital-motion decay consistent with the gravity-wave emission according to general relativity with an accuracy better than 0.5% (Taylor et al. 1992). This timely result would provide further impulse to ongoing discussions about plans to build large-scale ground-based laser interferometers. Their large bandwidth would allow detection of gravitational waves from a very wide range of potential sources. The following year, the 1993 Nobel Prize in Physics was awarded to Russell A. Hulse and Joseph H. Taylor for the discovery of the binary pulsar PSR 1913+16. Subsequent observations and interpretations of the evolution of the orbit, had opened up "new possibilities for the study of gravitation."[141]

---

[139] See preface in (Hough et al. 1989).

[140] According to the Italians, "In spite of a few European meetings, and a good collaboration with German and British colleagues through two European grants, we were actually pushed in the direction of a bi-national project, rather than a joint two-detector European project, by the fact that the German team at Garching and the British team (mainly at Glasgow) were pushing their own national projects, and feared that the settlement of a European collaboration would delay their acceptation" (Bradaschia 2009, 6). Both the tension existing between national ambitions and efforts towards international collaboration, and the problem of an absence of a really coordinated gravitational-wave community at a European level, played a negative role in this phase. On the question of why European leading groups in the field of gravitational-wave interferometry did not join forces to build a European observatory with at least two detectors at kilometer scale see forthcoming contribution: Adele La Rana, The origins of Virgo and the emergence of the international gravitational wave community, in (Blum et al. 2020).

[141] The Nobel Prize in Physics 1993, NobelPrize.org. Nobel Media AB 2020,

However, more menacing clouds were gathering on the horizon for the planned British-German 3-km gravitational-wave antenna. In November 1989, the Berlin Wall had been opened after nearly three decades, marking the falling of the Iron Curtain, and in August 1990 the Reunification Treaty between the two German states was signed. In mid-1990 BMFT formed a multidisciplinary advisory commission led by the theoretical physicist Siegfried Großmann, which was supposed to make recommendations about fundamental research in Germany. The commission worked from August 1990 to November 1991 and a 124-page long report was officially released in April 1992. Notwithstanding the *special sympathy* with which the Commission regarded the large-scale experiment of a gravitational-wave detector "because of its novel scientific objectives", also acknowledging its "*special charm*" due to its innovative approach to gravitation, "the smallness, possibly the still-undetectability, of the effect," was also highlighted in the report.[142] The search for gravitational waves was not the only field affected by this report, and there were hints of disciplinary rivalries in the outcome: On one side, BESSY II, the upgraded new electron storage ring producing synchrotron radiation for materials research purposes to be built in Berlin (more on this later), was considered a "high priority" initiative, that already in July of that year the project got the "green light". On the contrary, fundamental, particle physics did not receive a favorable treatment, but instead a financial horizon which "should not enlarge, nor back off in the next few years". It was easy to imagine "what a hard standing the three solid-state physicists in the commission had, to enforce this formulation." As for gravitational waves, the commission had not recommended "immediate implementation, but swift prosecution with intensive scientific discussion" (Dreisigacker 1992, 374).

Thus, also endorsed by the Großmann commission's report, BMFT took a position fully justified by the critical situation due to German reunification and the challenging responsibility in the process of restructuring East German science (Sabel 1993): the ambitious dream of a 3-km

---

https://www.nobelprize.org/prizes/physics/1993/summary/, accessed 1/02/2020.

[142] The Commission remarked that the "big requirements for extreme stabilization of lasers, mirror technology are promising high technical spin-off." It was also reported that the 1987 DFG *Denkschrift* on Astronomy (Völk et al. 1987) had recommended the gravitational-wave detector. A copy of the pages in the commission's report related to the project of building a detector for gravitational wave astronomy (pp. 76-78) can be found in the Archives of the Max Planck Society in Munich (ARMPG, Fol. 373-380).

interferometer was definitely not to be considered a priority in respect to other planned physics projects, on which BMFT had started huge investments programs since the mid-1980s.[143]

The Max Planck Society had officially asked for support for the big project by end 1989/early 1990.[144] However, it became evident as time passed, that there would be no action on the side of the BMFT, notwithstanding many years of financial support for in-depth and outstanding preliminary investigations. Initially, during 1991, it had become already clear that BMFT would promote the project *only* in connection with a strong European cooperation.[145] However, BMFT was "still very undecided" and once the British were putting everything on hold for financial reasons, Max Planck people hoped to counteract such hesitating attitude with a multinational initiative involving the French and the Italians.[146] The project had actually prominent and constant support within the Society by its Vice-President Herbert Walther, Director of the MPI of Quantum Optics, who had asked Danzmann to come back from Stanford University and lead the gravitational-wave group. There was even an attempt to get support from Walter E. Massey, Director of the National Science Foundation. In April 1992, at the time of the official release of the Großmann commission's report, Massey sent a letter to the Federal Minister Heinz Riesenhuber, describing the crucial importance of such a network of detectors, as "one of the outstanding opportunities in experimental science today," especially emphasizing how one of the important criteria used in their evaluation had been "optimal performance with a possible European detector site, assumed to be in Hannover, Germany." The answer came only next September, after several months: "As a number of proposals had been submitted regarding investment in basic research, a decision required detailed evaluation. Careful consideration of the priorities of basic scientific research has revealed that BMFT funding of the gravitational wave detector will not be possible in the foreseeable future. I regret that this project, like other projects concerned with interesting scientific topics, cannot be supported by BMFT."[147] This answer left little room for doubt on BMFT's intentions. However, it looked like the uncertainty in BMFT was arising from different interpretations of the Minister's directive: Not to start any new "big project".[148] Soon after, in a letter to Edmund Marsch, deputy Secretary General of the Max Planck Society, Danzmann, leader of the project, stressed that "The Department [for Basic Research] is interpreting this decision in such a way that it is no longer possible to promote basic research in these areas within the framework of 'Verbundforschung'." For this reason, their new application was not allocated federal funds, "despite the opinion of the evaluation committee".[149] After a few days Danzmann wrote again to Marsch: "We understand that in the current financial situation, the BMFT is unable to make a capital cost contribution to the construction of a gravitational wave detector."[150]

---

[143] As recalled by Hermann Schunck (at the time director at BMFT and responsible for fundamental research especially for Physics) the Ministry neither got any budget hike to take up that new responsibility nor any extra personnel. He further emphasized: "But there was another reason of psychological importance. BMFT had started a huge investment program in physics in the middle of the 1980s, including building a huge HEP-accelerator (HERA) in Hamburg (with prominent participation of Italy), an enlargement of experimental possibilities for nuclear physics in Darmstadt, a new nuclear research reactor in Berlin, an X-ray satellite for the Max Planck Society and some more. All these projects proved to be highly successful, by the way. The political leadership of BMFT had the clear feeling that all this was enough for physics." Hermann Schunck: Written interview by Adele La Rana, May 14, 2019.

[144] See related documents in AMPG, II Abt., Rep. 66, No. 3853, 65.

[145] See for example memorandum dated December 6, 1991 written by Daniel Cribier, head of the French project, following a meeting in Munich with Wolfgang Hasenclever, General Secretary of the Max Planck Society, Walther, Director of the Institute for Quantum Optics, and Danzmann, now leading the GEO project (ARMPG, Fol. 382).

[146] Hasenclever to Dieter Kind, President of the Physikalisch-Technisch Bundesanstalt, August 14, 1991, ARMPG, Fol. 387.

What looked like the last word on the question was written on November 17, when the state secretary Gebhard Ziller answered to Wolfgang Hasenclever, General Secretary of the MPG. The latter had underlined how such big device was only aiming at fundamental research, but Ziller sharply answered that there would be priorities in the coming years and he very much regretted that he could not currently envisage further funding for the GEO project, but he hoped that the money spent so far had created the base on which MPG could and would continue to

---

[147] Walter E. Massey to Heinz Riesenhuber, April 28, 1992; Riesenhuber to Massey, September 2, 1992, ARMPG, Fol. 311 and 301.

[148] Since the late 1980s, there had been a shift in German Federal research policy regarding the responsibilities of the Max Planck Society. In previous decades, it had often been the case that the MPG had taken over the stewardship of large infrastructural projects, including ground-based astronomical observatories and satellites. This had put the MPS at an advantage with respect to other national institutions, which increasingly protested their dominant position. New initiatives such as the Denkschrift of 1987 called for a more horizontal distribution of tasks, leading to what was called "Verbundforschung", a form of organization that coordinated all German partners participating in large international projects. The gravitational wave interferometer had already been assigned funds under this scheme, making it very difficult to deny that a "large project" was on the horizon. For more on details on Verbundforschung see our upcoming book on the history of astronomy and astrophysics in the Max Planck Society.

[149] Danzmann to Marsch, October 14, 1992, ARMPG, Fol. 298.

[150] Danzmann to Marsch, October 19, 1992, ARMPG, Fol. 299. Our translation.

work to some extent on the issue of gravitational waves using its own funds for fundamental research.[151]

After the detection of gravitational waves in 2015, it became clear in retrospect that this was a missed opportunity, as described, for instance, in a public interview by Danzmann himself. He stated that in Germany the wrong decision was made: while the Germans developed much of the high technology for the detection, the Americans by investing in the buildings and vacuum chambers of the large detectors claimed the largest recognition (Hilbig et al. 2017).

The most dramatic account and interpretation of this retreat from a full-sized interferometer is given by Hermann Schunck, at the time ministerial director at BMFT and responsible for fundamental research especially for Physics:

This situation was a catastrophe for pledging for a new project, what my unit did after the letter had arrived. While discussing the project with our superiors I was bluntly told to take my hands off that subject altogether. This was a situation quite new and exceptional for me. I was used to be able to do my job basically on my own responsibility and judgment, within certain constraints, like budget and general policy, and with the general hierarchal decision process of a political bureaucracy. My immediate judgment what this drawback meant to GW research was: not just a usual missed opportunity like any other, but a missed star hour of the history of Physics in Germany. I understood the importance of GW as the last great prognosis stemming from General Relativity that had not been proven experimentally. And I had the clear vision that the first group to do this would travel to Stockholm. Working as a research administrator you do not have many chances like that, if any.[152]

As historians, however, we must also consider that German Unification had changed circumstances in a truly dramatic fashion: By early 1993, it had become definitely clear that the British Science and Engineering Research Council would not fund the joint project "for financial reasons." Even in the United States, which inevitably served as a reference for investment in large research projects, the early 1990s was a particularly difficult period for physics, and one which for the first time featured an open conflict between different branches (particle-, solid state- and astrophysics) for a reducing pot of resources: the Superconducting

---

[151] Hasenclever to Ziller, November 4, 1992, ARMPG, Fol. 281; Ziller to Hasenclaver, November 17, 1992, ARMPG, Fol. 279-280.

[152] Hermann Schunck: Written interview by Adele La Rana, May 14, 2019.

Super Collider was cancelled in 1993 (Riordan et al. 2015; Martin 2018). Within astrophysics itself, gravitational wave detection was perceived as competing with other projects in the field: the 1990s decadal survey, which guides American investment in astronomy and astrophysics (National Research Council 1991), was negative to gravitational waves, and the community was particularly allergic to the claim of efforts in the field aiming towards an "observatory" rather than a high-risk detection experiment. With conflicting interests on a "knife-edge", what ultimately saved LIGO was "pork-barrel" federal politics in the United States, which favored investment in the geographical sites in Washington State and Louisiana (Collins 2004, pp. 489-511).[153] While in Germany the scientific communities were similarly split, the geographical-political circumstances following unification were much more disfavorable to gravitational waves: while there was strong regional interest in Bavaria and especially Lower Saxony, German federal research priorities were completely oriented towards areas formerly within East Germany. BESSY II (mentioned earlier) was ultimately pursued because it would be located in East Berlin; and during the same period, the flagship institute of the Garching area, the Max Planck Institute for Plasma Physics, was forced to focus its further expansion in Greifswald, on the Eastern German Baltic Sea coast.

## 11. GEO600: A retreat from full-scale experimentation to focus on instrumental developments

After all those efforts, with ongoing plans for similar large-scale antennas both in Europe and the U.S., the German and British teams who had pioneered research in the field since the beginning of the 1970s, and had a longstanding collaboration being both active building prototype interferometers, were deeply disappointed and felt they could not renounce. They struggled to find an alternative strategy, as pursued by Karsten Danzmann, who by the time of the failure had already been appointed as professor in Hannover with the perspective of the full-scale experimental enterprise, before it was known that even the funding from the state of Lower Saxony had been reallocated.[154]

---

[153] After listing some potential causes for the final decision ("skill at lobbying of the various parties", Washington's need of big science because of its prestige, or because "a congressional staffer explained that the cost of LIGO was only an accounting error in the size of budget they were dealing with") Collins remarks, "So why was LIGO funded? […] The sociological interest is that the forces of all kinds were roughly evenly balanced, so the funding decision could have gone either way […] The funding of LIGO was an immensely important issue to the scientific community, which works with the NSF, but it was not an immensely important issue to those who were actually providing the money. None of the energy and hard work that both the pro-LIGO side and the anti-LIGO side put into presenting their cases was wasted, but the net result was an even balance […] So, there is no big story about the funding of LIGO except the story that put the funding on a knife edge." (Collins 2004, 509-510).

The arm length, a most important design parameter, turns out to be the major cost factor: the cost of civil engineering and of the vacuum system being approximately proportional to the length, making up close to 70% of the total cost. Thus, a reduction in arm length would cut down the detector cost considerably, making the plan to build a much smaller facility a realistic aim for the British-German teams. Max Planck scientists thus joined forces with British researchers to build the smaller GEO600 "experiment", a 600-m arm-length antenna (Danzmann et al. 1994; Lück and the Geo600 Team 1997).[155] GEO600, whose construction began in September 1995, was designed on the basis of experience with two prototypes (in view of an interferometric detector with arms of a length close to 3 kilometers): the 10-meter interferometer at the University of Glasgow and the 30-meter interferometer at the Max Planck Institute of Quantum Optics. It had been decided that it should be something built "with their own hands", which could give them a chance to develop new technology, "piece by piece", and get money for each innovative "standalone project" from different sources, even from the BMFT. Crucially, the raised money was never explicitly for the *detector* itself, it was always for specific technology developments.[156]

In 1993 Danzmann had become professor at the University of Hannover and Director of the Institute for Atomic and Molecular physics, and at the same time, since 1994 he was also made leader of the new branch of the Max Planck Institute for Quantum Optics in Hannover.[157] This consolidated the leadership of the German side of the collaboration under the same person.

---

[154] Walther to Hasenclever, April 24, 1992, ARMPG Fol. 271-2.

[155] In the meantime the whole funding structure had changed in the U.K.: the new Council was in favor of the gravitational wave project (Bernard Schutz: Interview by Adele La Rana, CERN, August 28, 2017), and so, GEO600 was funded by the Max Planck Society, the Volkswagen Foundation, the State of Lower Saxony, and by the Particle Physics and Astronomy Research Council on the British side. The project was soon presented at the first Edoardo Amaldi Conference held in Frascati, Italy, June 14-17, 1994 (Danzmann et al. 1995). On GEO600, see also (Völter 2016).

[156] During the year 1993-2000, this strategy was supported by Hermann Schunck at BMFT, who was able to channel "leftover" money from other German projects, that had not been able to spend it, for financing specific GEO600 needs justifiable as "standalone projects" such as vibration isolation, data acquisition, novel optics, laser stabilization, novel vacuum system design etc. (Hermann Schunck: Written interview by Adele La Rana, May 14, 2019, and personal communication to the authors, November 20, 2019).

[157] ARMPG, Fol. 246, 212-213.

Despite the drawbacks, gravitational waves were now becoming a most important new field of research and one that spanned across several branches of the Munich 'family' of institutes including Astrophysics, Quantum Optics, and, later, the two sites of Gravitational Physics in Golm and Hannover. In March 1991, the Max Planck Institute for Physics and Astrophysics had been split up into three independent institutes: the MPI for Physics in Munich (Werner-Heisenberg-Institut), the MPI for Astrophysics and the MPI for Extraterrestrial Physics, the last two in Garching.[158]

In 1990, the Max Planck Society was in involved in supporting the state of Brandenburg in the plan to found a large "Blue-List" institute as successor of the Zentralinstitut für Astrophysik (ZIAP) of the former GDR's Academy of Sciences. In order to counteract the closure of that GDR institute as recommended by the German Council of Science and Humanities (*Wissenschaftsrat*), and in particular of the Einstein-Laboratorium for Theoretical Physics in Potsdam, which was part of it, Jürgen Ehlers took over the initiative to propose the creation of a Max Planck Institute for Gravitational Physics in Potsdam.[159] This was justified in particular by the historical connections of Einstein to this city (Hillebrandt 2013; Goenner 2016, 2017).

---

[158] Minutes of the 127th Senate meeting in Frankfurt am Main, 08.03.1991, AMPG, II. Abt., Rep 60, No. 127.SP, pp. 23-24.

[159] A memorandum for the reorganization of the Einstein-Laboratorium in Potsdam into an International Einstein Center had already been formulated by Hubert Goenner and Friedrich Hehl in February 1991 and submitted to the secretary of the German Council of Science; soon after it was announced in (Goenner and Hehl 1991). The involvement of the Max Planck Society and the evolution into the decision to found a dedicated institute for gravitational physics was summarized in a letter from President Zacher to the influential relativist Peter Bergmann, who had written an open letter circulated to other well-known colleagues in the field, expressing their concern for the further maintenance of the Einstein Laboratory in Potsdam (Zacher to Bergmann, December 19, 1991, AMPG, II. Abt. Rep. 62, No. 205, Fol. 198). In July 1990, the Government of the GDR and the Federal and State Governments authorized the Science Council (*Wissenschaftsrat*) to evaluate the research institutions (with the exception of the universities) of the GDR. In the unification treaty this assignment was confirmed and extended by the request that recommendations be formulated concerning the future of these institutions in the enlarged Federal Republic of Germany. In September 1990, a committee of physicists visited several institutes of the former GDR, among them the Einstein Laboratory in Potsdam. Based on the report of this committee and on further discussions with scientists, the Science Council recommended in its final report, published in July 1991, that the Einstein Laboratory be discontinued. At the same time, in recognizing that research on gravitation was severely unrepresented in Germany, the Council recommended to found an "Albert Einstein Institute for Gravitational Physics in the Berlin-Potsdam area. In the meantime, in accordance with such recommendation and on the initiative of the president Zacher, a committee was formed to elaborate a memorandum concerning the possible scientific and organizational structure of such an institute, which was supposed to carry out research and teaching in modern areas of the theory of gravitation, particularly its relation to quantum field theory. See general discussions on the ZIAP question in 1991 at the October meeting of the presidential committee of the Max Planck Society, chaired by Heinrich Völk and formed by President Hans Zacher at the time of reunification, in order to answer a list of specific questions and to give recommendations on further developments of astronomy and astrophysics, indicating especially promising areas for basic research—which is typically to be conducted at Max Planck Institutes—and to set priorities (AMPG, II. Abt., Rep. 62, No. 17, Fol. 431-433. Materials about the activity of the presidential commission can be found in Rep. 62, No. 17). A chronology of the steps leading to the

In 1995, the Max Planck Institute for Gravitational Physics named after Albert Einstein, the physicist who developed the theory of general relativity (Albert-Einstein-Institut, AEI), was founded in Golm, near Potsdam, with Directors Jürgen Ehlers and Bernard F. Schutz, who also remained part-time in Cardiff.[160] Immediately after the foundation of the institute, Hermann Nicolai was appointed as third director at AEI.[161] The creation of a new astrophysics-oriented Max Planck Institute in the new *Bundesländer* following German unification, resulted from a further "cell division" in the Munich area (Trümper 2004) and increased the dominance of the Max Planck Society in the astronomical-astrophysical research fields.

The research program had its roots in several activities already ongoing during the previous 25 years at the Institute for Physics and Astrophysics: the foundations of general relativity, the quest for unifying general relativity and quantum mechanics, the study of neutron stars and black holes, and, of course, gravitational wave antennae. In 1995, in parallel with the founding of the Albert Einstein Institute, the construction of the 600-meter arm-length detector was starting in Ruthe, a site 20 km south of Hannover. Methods for data analysis and simulations of possible sources were developed both at the University of Wales in Cardiff and at the Albert Einstein Institute in Potsdam. Soon after, this activity became one of the main research focuses

---

formation of a commission for the founding of an "Albert-Einstein-Institut für Gravitationsphysik" Max Planck Institute for Gravitational Physics is outlined at the beginning of the minutes of the meeting of October 19, 1993, also containing Ehlers' proposal and memorandum (AMPG, II. Abt., Rep. 62, No. 205, Fol. 6-20).

[160] The foundation of the Max Planck Institute for Gravitational Physics in the context of German unification was discussed in a dedicated committee (*Beratung über Aufnahme von Forschungsaktivitäten der MPG in den neuen Bundesländer nach der Vereinigung*) as of October 1990 (see CPTS meeting minutes of 2.10.1990, including a long report by the MPG President Hans F. Zacher, and minutes of 07.02.1991, 05.06.1991, 23.10.1991, 07.02.1992, 19.10.1993, AMPG, II. Abt., Rep. 62, No. 1821, 1822, 1823, 1824, 1825, 1830). In October 1991, Ehlers had in fact prepared a memorandum about his plans for such an institute that he had sent to Zacher (see a copy of this document within the CPTS meeting minutes of 16.10.1992, AMPG, II. Abt., Rep. 62, No. 1827). A special committee for the foundation of a Max Planck Institute for Gravitational Physics was formed on 7 February 1992. Works of the committee were reported during several meetings of the CPT Section (CPTS meeting minutes of 07.02.1992, 03.06.1992, 16.10.1992, 03.02.1993, 16.06.1993, 19.10.1993, 03.02.1994, 08.06.1994, 09.02.1995, AMPG, II. Abt., Rep. 62, No. 1825, 1826, 1827, 1828, 1829, 1830, 1831, 1832, 1834). Details about Ehlers' role and other aspects related to the phase preceding the actual decision to found the new institute—also connected to the above-mentioned difficult period of the Institute for Astrophysics following Kippenhahn's anticipated retirement—can be found in (Hillebrandt 2013). For a related discussion on the German unification phase, see (Dreisigacker 1991). On the foundation of AEI, see also (Goenner and Hehl 1991; Goenner 2016). For the evolution of research on general relativity in Germany and its eventual institutionalization in the form of a dedicated research institute, see (Goenner 2017a).

[161] CPTS meeting minutes of 09.02.1995, 21.06.1995, 19/20.10.1995, 8/9.02.1996, AMPG, II. Abt., Rep. 62, No. 1834, 1835, 1836, 1837. Ehlers, Schutz and Nicolai led research activities respectively in general relativity, relativistic astrophysics and quantum gravity/unified theories. Numerical relativity and computer simulations, also related to collapsing relativistic binaries and their associated gravitational waves, were an active part of the research activity since the very beginning of AEI, see for example (Schutz 1999).

of the new Institute with the decision to transform the already existing research center at the Max Planck Institute of Quantum Optics, based in Hannover and led by Karsten Danzmann, into a branch of the Albert Einstein Institute. The founding of a "center of excellence" for gravitational wave research thus unified both experimental and theoretical activities under the same roof.[162]

In 2001, Danzmann was promoted to Director of the Laser Interferometry and Gravitational Wave Astronomy Division, the first of two divisions which were planned when the Quantum Optics branch in Hannover became officially part of the Albert Einstein Institute, which has since then sites in both Potsdam and Hannover.[163]

The move to Hannover itself, where Danzmann spent his early career, reflected the politics of increasingly large projects in gravitational wave detection; while researchers at Garching had attained worldwide recognition for their development of multiple ingenious experimental methods for improving the sensitivity of the detection of the space disturbance, two other crucial factors were outside Bavarian control: the laser source itself and the site for the large interferometer. The laser development efforts were being concentrated in Hannover, in the group of Danzmann's scientific mentor Herbert Welling, and through his worldwide prominence in the field, Welling obtained the siting of the large interferometer for the state of Lower Saxony,[164] also facilitated by the promise of co-funding from the nearby Volkswagen Foundation (Grote 2018, 72–77).

---

[162] In June 2000, the founding of a center for gravitational wave research was discussed during a meeting of the CPT Section. As stressed by Bernard Schutz, the whole operation would assure participation of the Max Planck Society with a cutting-edge role in the outstanding projects EURO and the laser-interferometric detectors LIGO and LISA, the Laser Interferometer Space Antenna mission, a giant interferometer to be placed in space. A committee was formed to examine the whole plan (CPTS meeting minutes of 07.06.2000, 19/20.10.2000, 15/16.02.2001, AMPG, II. Abt., Rep. 62, No. 1851, 1852, 1853).

[163] CPTS meeting minutes of 15/16.02.2001, 20.06.2001, 18/19.10.2001, AMPG, II. Abt., Rep. 62, No. 1853, 1854, 1855.

[164] Interview with Karsten Danzmann, March 29, 2018, Deutsche Physikalische Gesellschaft e. V., Stern-Gerlach-Medaille 2018, available at https://www.youtube.com/watch?v=tNTB74bFGuc, accessed 23/2/2020. Danzmann is considered part of the so-called "Welling Laser Family" (*Laserfamilie Welling*), and just a few years before, Welling had consolidated the region's footprint in this field with the establishment of the Laser Zentrum Hannover (Liftin and Mlynek 2009). For the latest account of his career, see: "Grosses Verdienstkreuz für Professor Herbert Welling", Presseinformation des Niedersächsischen Ministeriums für Wissenschaft, 31.8.2019, available online at https://www.mwk.niedersachsen.de/startseite/aktuelles/presseinformationen/grosses-verdienstkreuz-fur-professor-dr-herbert-welling-180202.html, accessed 23/2/2020.

Interestingly, this geographical move did not imply the geographical relocation of personnel, as it coincided with a most radical stage of generational renewal occurring during the 1990s: the experts from Bavaria from the founder years gradually transferred their technologies and practical know-how to younger members hired directly in Hannover, starting with Danzmann himself, who did not know much about the subject when he was first hired to lead the project. Over the course of the decade, as Bavarians reached the end of their careers, the positions freed by them were used to hire a new generation in Lower Saxony.

While the German 3-kilometer interferometer project had to be put aside in favor of the smaller GEO600, the American proposal for the Laser Interferometer Gravitational-Wave Observatory (LIGO), consisting in two widely separated long-based installations (4 km arms) within the United States, was funded, like the Italian Virgo.[165] The Virgo project for a 3-km interferometer was approved between 1992 and 1994 by the Centre National de la Recherche Scientifique (CNRS) and the Istituto Nazionale di Fisica Nucleare (INFN), eventually leading to the building of the Virgo interferometer at Cascina, near Pisa, beginning in the second half of the 1990s.

In 1997, the British-German collaboration finally entered in partnership with LIGO, becoming part of the worldwide network of gravitational wave detectors and contributing to the next generation of US detectors with new advanced technologies (B. Abbott et al. 2004; Dooley et al. 2016).[166] A collaboration linking the LIGO detectors in the U.S. with its partners GEO600 in Germany and the Virgo detector in Italy was established in early 2007. Many of the technologies developed at GEO600 became thus instrumental in enabling the unprecedented sensitivity of LIGO and Virgo.[167]

---

[165] The LIGO construction proposal was approved by the National Science Board in 1990, and in 1992 the LIGO cooperative agreement for the management of LIGO was signed by NSF and Caltech, while construction at the chosen sites Hanford and Livingston began between 1994 and 1995.

[166] Since 2001, when the Hannover branch of the Max Planck Institute of Quantum Optics merged with the Albert Einstein Institute, GEO600 has been operated by AEI within the international collaboration with the Leibniz University of Hannover (which had been actively involved in the program through Karsten Danzmann) and the University of Glasgow and University of Wales at Cardiff and is now part of the worldwide network of gravitational wave detectors, including LIGO in the U.S., Virgo in Italy, and KAGRA (Kamioka Gravitational Wave Detector) in Japan, which has been completed in October 2019. With two 3-km baseline arms stretching through tunnels under a mountain, it is the world's first interferometer of its size to be built underground. For a list of institutions and members of the LIGO Scientific Collaboration, see (LIGO Scientific Collaboration 2018).

[167] We cite here a series of articles related to the first realization by the German group of innovative detector technologies which made Advanced LIGO and Virgo so sensitive, such as the resonant sideband extraction, the automatic alignment as well as the power and dual recycling in suspended interferometer, the thermally adaptive optics, the detuned dual recycling, the contactless mirror refocusing, the automatic beam alignment, the radiation pressure interferometer calibration, the stable high-power lasers, the DC readout of signal recycled interferometer,

At the same time, the European gravitational wave community joined in the ESA-NASA space project LISA, the Laser Interferometer Space Antenna, consisting of three spacecraft in heliocentric orbits, forming an equilateral triangle with 2,5-million-km sides, which would be complementary to ground-based detectors making it possible to detect the extremely low frequency ranges—and thus also different kind of sources—which are limited by noise of different origins (mainly seismic and Newtonian) affecting interferometric gravitational wave detectors located on the Earth's surface (Danzmann et al. 1993; Rüdiger et al. 2001). LISA is a very long-term project, which at the time was expected to measure gravitational waves only several decades after its inception.

## 12. Open questions: A Munich 'family affair' of theory and instrumental specialization in a global scientific race?

On 14 September 2015, at 09:50:45 UTC, 100 years after Einstein formulated the field equations of general relativity, the two detectors of the Laser Interferometer Gravitational-Wave Observatory (LIGO) simultaneously observed a transient gravitational wave signal matching the waveform predicted by general relativity for the inspiral and merger of a pair of black holes of about 30 solar masses each. The signal caused the mirrors at the ends of each interferometer's 4 km arms to oscillate with an amplitude of about $10^{-18}$ m, roughly a factor of a thousand smaller than the classical proton radius. It was the first direct detection of gravitational waves after decades of experimental efforts and the first ever observation of a binary black hole merger (LIGO Scientific Collaboration & Virgo Collaboration et al. 2016), a culminating achievement in the long process of the renaissance of general relativity.

Events such as the first one detected by the LIGO collaboration, which was given the name GW150914, are invisible for traditional astronomical instruments, as any signal other than gravitational waves is emitted near the merging black holes. But then, on 17 August 2017, four decades after Hulse and Taylor discovered the first neutron star binary, the Advanced LIGO and Advanced Virgo observatories made their first direct detection of a swell of gravitational

---

the contactless mirror refiguring and the first realization and routine operation of squeezed light in a large gravitational wave detector (Mizuno et al. 1993; Heinzel et al. 1996; Schnier et al. 1997; Heinzel et al. 1998; Heinzel et al. 1999; Lück et al. 2000; Freise et al. 2000; Heinzel et al. 2002; Lück et al. 2004; Grote et al. 2004; Mossavi et al. 2006; Seifert et al. 2006; Hild et al. 2009; Grote 2013; Wittel et al. 2014).

waves from the coalescence of a neutron star binary system, which was followed after 1.7 seconds by a burst of gamma rays detected by the orbiting Fermi Gamma-Ray Space Telescope and INTEGRAL observatory (LIGO Scientific Collaboration & Virgo Collaboration 2017). The detection of this new gravitational-wave signal (GW170817) offered a novel opportunity to directly probe the properties of matter at the extreme conditions found in the interior of these stars, while the unprecedented joint gravitational and electromagnetic observation of this astronomical cataclysm was marking the beginning of a new era in multi-messenger astrophysics (Abbott et al. 2017a, 2017b).[168]

Violent astrophysical events where large masses undergo large accelerations (such as gravitational collapse and merging of compact objects) can be very powerful sources of gravitational waves, a messenger of the high-energy universe together with cosmic rays, gamma rays and neutrinos (Lipari 2017). The next step might be the associated detection of high-energy neutrinos from binary neutron star or black hole mergers which are primary sources of gravitational waves. Given the sensitivity of the gravitational-wave signal to the neutron star structure, the new era of multi-messenger astronomy, with its growing synergy between astrophysics, gravitational physics and nuclear physics, will also provide new insights into the nature of dense matter and into the properties of new states of matter at exceedingly high density and temperature.

Many of the instrumental innovations that eventually led to the first 2015 detection of gravitational waves using the LIGO detectors had been pioneered by Max Planck Institute researchers,[169] and they also played a key role in the computational tasks related to the detection efforts.[170] The LIGO Scientific Collaboration also includes two Russian groups from Lomonosov Moscow State University and from the Institute of Applied Physics, Russian Academy of Sciences.[171]

---

[168] The paper (Abbott et al. 2017b) describing the multi-messenger observations was co-authored by almost 4000 scientists from more than 900 international institutions, using 70 ground- and space-based observatories. Electromagnetic observations revealed signatures of recently synthetized material, including elements such as gold and platinum, showing that the rapid neutron-capture process needed to build up many of the elements heavier than iron might take place primarily in neutron-star mergers, and not in supernova explosions. A great number of papers on the different observations appeared in the same issue of the *Astrophysical Journal* (Vol. 848, No. 2, 2017).

[169] The Hannover/GEO principal contributions to LIGO include the following: The laser system; The demonstration of squeezed light at GEO600 and development of novel ideas to control squeezed light in GW detectors; Techniques for lock acquisition and for alignment control of mirrors such as beam centering on wavefront sensors; The demonstration of several technologies in GEO600, which lead to their adaptation in Advanced LIGO: signal recycling, electrostatic actuators, multi-stage mirror suspensions with monolithic last stage, squeezing application, thermal compensation systems to shape mirror geometries. The main Hannover/GEO contributions to Virgo include: Parts of the laser technology; Beam centering technology for the automatic alignment system of mirrors; The squeezed light source and corresponding support in operation at Virgo today.

After such breakthroughs events, an open question already addressed by the gravitational wave community became more pressing: what were the reasons for the failure of early attempts at an extended European collaboration aiming at a ground-based twin interferometer project including the French-Italian-born Virgo? (La Rana and Milano 2017, 194).[172]

This very preliminary outline of some main aspects of the story behind gravitational waves has helped shed some light on to what extent further wider and in-depth historical studies are needed to reconstruct the dynamics of such a missed opportunity to realize a European network of gravitational-wave telescopes that might have followed and matched the successful example of effective cooperation in the CERN enterprise.

---

[170] The development of highly accurate analytical and numerical models of gravitational-wave sources—in particular of gravitational waves that neutron stars or black holes generate in the final process of orbiting and colliding with each other—have allowed extraction of astrophysical and cosmological information from the observed waveforms. These waveform models are then implemented and employed in the continuing search for binary coalescences. To significantly increase the probability of identifying gravitational waves in LIGO and Virgo data, the search for burst-like events in turn requires detailed knowledge of the expected signals from different sources and such search tools are sensitive because of systematic development in the algorithm and methods. Numerical relativity simulations with supercomputers not only play an important role in predicting gravitational waveforms that are used for gravitational wave detection, but allow in general exploration of general relativistic phenomena and other high-energy phenomena, such as gamma-ray bursts and stellar core collapse, or mass ejection with related nucleosynthesis processes.

[171] Both groups have been responsible for separate functional units of the LIGO detectors. Research performed by the Braginsky group at the Physics Department of Moscow State University since the early 1990s also made a significant contribution to the development and fabrication of LIGO gravitational wave detectors. After first starting research in the field of gravitational physics in the early 1970s, Braginsky's group began to work on gravitational wave laser antennae in the early 1990s. The main aspects of this research are described in (Braginsky et al. 2016). On the contribution of the Applied Physics Institute, see (Khazanov and Sergeev 2017). For a summary of experimental research on the detection of gravitational radiation performed in the Soviet Union, see (Rudenko 2017).

[172] See especially La Rana's forthcoming contribution in (Blum et al. 2020).

**Acknowledgments**

We are very grateful to Walter Winkler and Hermann Schunck for some specific recollections, comments and extremely useful suggestions and to Karsten Danzmann and Hartmut Grote for providing extremely useful material and information related to the contributions of GEO600 to LIGO and Virgo.

Special thanks go to Adele La Rana for reading and commenting on a preliminary draft of this work, for a very productive exchange of ideas and in particular for having kindly given us access to some interviews she did with protagonists of these events.

We are also grateful to Heinrich Völk for his kind permission to cite an excerpt from our conversations in Heidelberg, as well as to Hermann Schunck and Bernard Schutz for giving us permission to quote parts of their interviews given to Adele La Rana.

Thanks are also due to the Niels Bohr Library & Archives, American Institute of Physics, for allowing permission to quote from the cited interviews. One of us (LB) is very grateful to Florian Spillert for his helpful assistance at the Archives of the Max Planck Society in Berlin.

The roots of this work date back to LB's participation in the international working group established at Department I of the Max Planck Institute for the History of Science, which started in 2014 in preparation of the centenary of Einstein's theory of general relativity, and continued in the context of our research on the history of astronomy, astrophysics and space sciences within the Research Program "History of the Max Planck Society". In this regard, we gratefully acknowledge discussions with members of the Program, who provided useful comments on a preliminary draft of this work.

Last but not least, our special gratitude goes to Roberto Lalli and Jürgen Renn for their attentive reading of our contribution, for their precious suggestions, remarks, and for enlightening discussions.